\documentclass[sigconf]{acmart}

\AtBeginDocument{%
  \providecommand\BibTeX{{%
    \normalfont B\kern-0.5em{\scshape i\kern-0.25em b}\kern-0.8em\TeX}}}

\setcopyright{acmcopyright}
\copyrightyear{2021}
\acmYear{2021}
\setcopyright{acmlicensed}\acmConference[ASSETS '21]{The 23rd International ACM SIGACCESS Conference on Computers and Accessibility}{October 18--22, 2021}{Virtual Event, USA}
\acmBooktitle{The 23rd International ACM SIGACCESS Conference on Computers and Accessibility (ASSETS '21), October 18--22, 2021, Virtual Event, USA}
\acmPrice{15.00}
\acmDOI{10.1145/3441852.3471208}
\acmISBN{978-1-4503-8306-6/21/10}

\usepackage{balance}       
\usepackage{graphics}      
\usepackage[T1]{fontenc}   
\usepackage{mathptmx}
\usepackage{color}
\usepackage{booktabs}
\usepackage{textcomp}
\usepackage{makecell}
\usepackage{xspace}
\usepackage{fancyvrb}

\usepackage{amsthm}
\usepackage{hyperref}
\usepackage{subcaption}

\usepackage{microtype}        
\usepackage{ccicons}          

\newcommand{\eg}{\textit{e.g.}\@\xspace}
\newcommand{\ie}{\textit{i.e.}\@\xspace}
\newcommand{\etal}{\textit{et al.}}

\usepackage{enumitem}
\setlist[description]{leftmargin=\parindent,labelindent=\parindent}

\def\plainkeywords{dataset; sharing practices; disability; repository; machine learning.}

\makeatletter
\def\@copyrightspace{\relax}
\makeatother
\begin{document}


\title{Sharing Practices for Datasets Related to Accessibility and Aging}

\author{Rie Kamikubo}
\affiliation{%
    \institution{College of Information Studies}
    \institution{University of Maryland, College Park}
    \streetaddress{4130 Campus Dr}
}
\email{rkamikub@umd.edu}
\author{Utkarsh Dwivedi}
\affiliation{%
    \institution{College of Information Studies}
    \institution{University of Maryland, College Park}
    \streetaddress{4130 Campus Dr}
}
\email{udwivedi@umd.edu}
\author{Hernisa Kacorri}
\affiliation{%
    \institution{College of Information Studies}
    \institution{University of Maryland, College Park}
    \streetaddress{4130 Campus Dr}
}
\email{hernisa@umd.edu}
  

\renewcommand{\shortauthors}{Kamikubo et al.}

\begin{abstract}
Datasets sourced from people with disabilities and older adults play an important role in innovation,
benchmarking, and mitigating bias for both assistive and inclusive AI-infused applications. However, they are scarce. We conduct a systematic review of 137 accessibility datasets manually located across different disciplines over the last 35 years. Our analysis highlights how researchers navigate tensions between benefits and risks in data collection and sharing. We uncover patterns in data collection purpose, terminology, sample size, data types, and data sharing practices across  communities of focus. We conclude by critically reflecting on challenges and opportunities related to locating and sharing accessibility datasets calling for technical, legal, and institutional privacy frameworks that are more attuned to concerns from these communities.
\end{abstract}



\maketitle
\section{Introduction}
\newcommand\datasetsTotal{137 }

\begin{figure}[t]
    \includegraphics[width=1\linewidth]{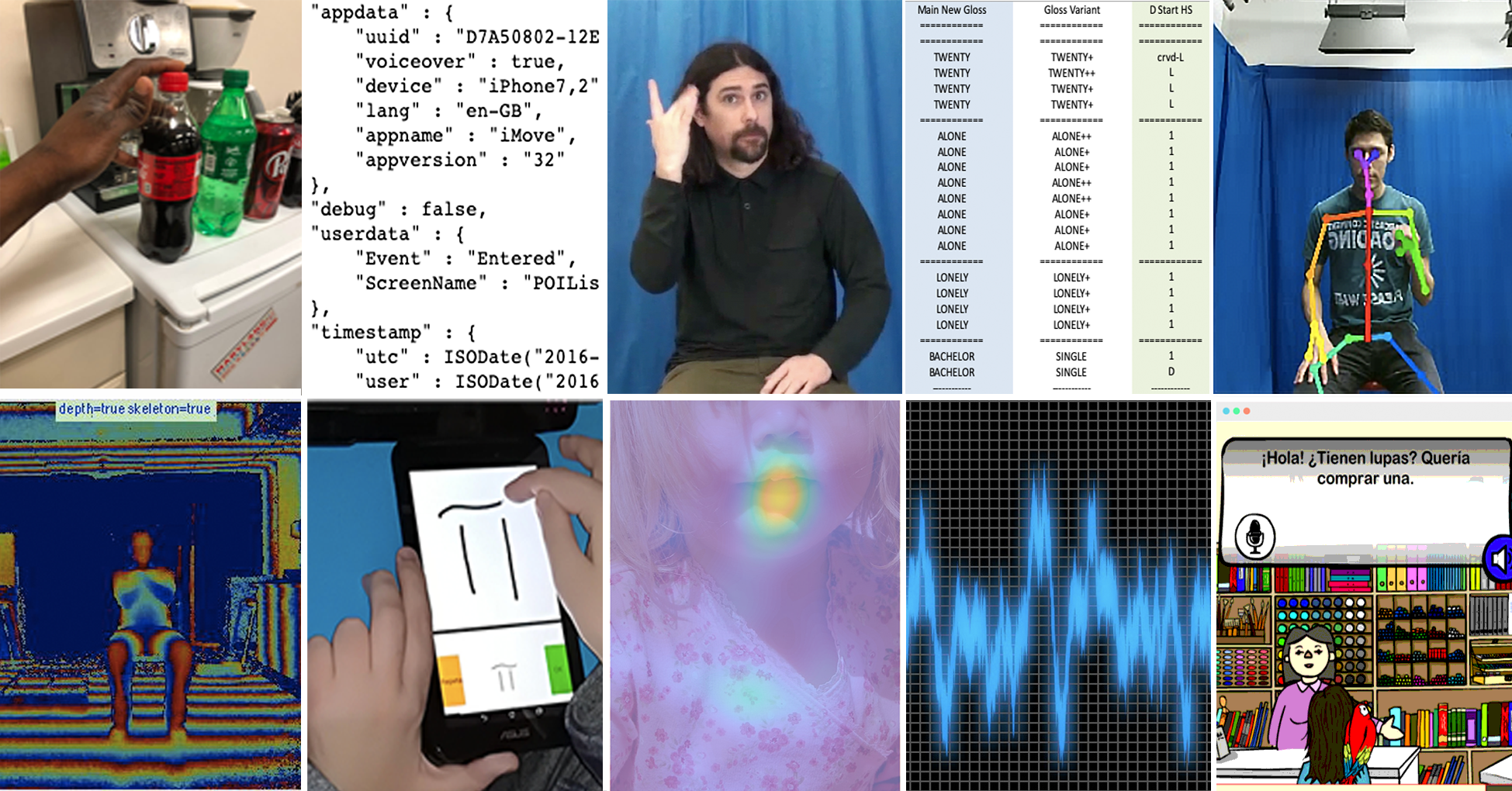}
    \centering
    \caption{Examples of accessibility datasets such as photos taken by blind users~\cite{lee2019hands}, assistive app logs of users with visual impairments~\cite{kacorri2016supporting}, sign language videos~\cite{huenerfauth2014release}, gloss annotations~\cite{neidle2012challenges}, motion captured signs~\cite{hassan2020isolated}, depth data from older adults' activities~\cite{leightley2015benchmarking}, stroke gestures by people with motor impairments~\cite{vatavu2019stroke}, eye-tracking data from autistic children~\cite{duan2019dataset}, voice recordings of people with speech impairments~\cite{cesari2018new}, and a speech corpus of people with intellectual disabilities~\cite{rello2014dyslist}.}
    \label{fig:example_data}
     \Description[10 screenshots from example datasets]{1. photo of soft drink bottles taken by a blind person with a hand in the frame on top of a bottle; 
     2. JSON formatted logs of mobile app usage from people with visual impairments,
     3. RGB video of a signer performing signs from a sign language dataset including 
     4. linguistic annotations shown as a table of terms and units for a sign language dataset,
     5. skeleton data from kinect sensor from a signer performing signs; 
     6. video frame of a 3D depth map data from Kinect sensor,
     7. finger drawn strokes on smart tablets from a person with motor impairments;
     8. saliency heatmap showing where autistic children are looking at in an image of a child,
     9. image of a sound wave collected from people with dysphonia, and finally,
     10. a screenshot of game where children with dyslexia perform conversational tasks of buying things from a shop}
 \end{figure}

A wide range of academic disciplines and industries source datasets from people with disabilities and older adults – each with its own data collection method as well as reporting and sharing practices.  A recurring challenge they face is navigating the tension between making their data accessible and restricting access to protect those represented in the data~\cite{hamidi2018should, sarwate2013signal, xafis2019ethics}.  This tension, typically faced by many social data stewards~\cite{lundberg2019privacy}, is not unique to accessibility.  However, there are benefit and risk traits unique to many accessibility datasets that can make the challenge more pressing, especially when they can be used in AI-infused applications.  

\textbf{Benefits.} Curation and sharing of accessibility datasets is crucial for innovation, benchmarking, bias mitigation, and understanding of real-word scenarios, where assistive and general purpose AI-infused applications are deployed.  For example, innovation in sign language processing requires videos sourced from the Deaf community~\cite{bragg2019sign, bragg2020exploring, lu2012cuny}.  Benchmarking visual question answering can benefit from real-world visual questions from blind people~\cite{gurari2018vizwiz}.  Bias mitigation could utilize features explicitly representing disability~\cite{trewin2019considerations}.  Understanding the use of mobility applications among blind people in the real world requires longitudinal-data from the intended users~\cite{kacorri2016supporting}.  More importantly, representation of people with disabilities and older adults in AI training datasets can contribute to more inclusive AI-infused applications~\cite{whittaker2019disability, trewin2019considerations, park2021understanding, guo2019toward}. 

\textbf{Challenges.} Despite their critical role, large datasets directly sourced from people with disabilities and older adults are scarce~\cite{kacorri2017teachable, bragg2019sign, morris2020ai}, either publicly available or not~\cite{kacorri2020incluset}.  This is partly due to smaller populations~\cite{sears2011representing}.  Other factors relate to high variability within a given disability or age group; data often being constrained to very specific tasks, applications, or scenarios making it difficult to aggregate; and annotations often requiring domain knowledge making it difficult to elicit through crowdsourcing tasks.  For instance, creating annotated video datasets for sign language synthesis requires linguistic background and sign language fluency~\cite{neidle2012challenges, kacorri2016data}. However, increasing representation of these populations yet can amplify ethical and privacy concerns derived from potential data abuse and misuse or re-identification risk~\cite{hamidi2018should,treviranus2019value,guo2019toward, abbott2019local}.  

In this paper, we explore how data stewards across different disciplines and over time have balanced risks and benefits when collecting, reporting, and sharing accessibility data. Specifically, we conduct a systematic review of \datasetsTotal accessibility datasets from 1984 to 2020.  As shown in Figure~\ref{fig:example_data}, we focus on annotated data resources generated from communities of interest in accessibility and aging that could be used to train or evaluate machine learning models.  Datasets are manually located over the period of two years and coded in terms of practices for reporting the communities represented;
sample size, data type, and how they relate to the purpose and tasks of the data collection; and sharing practices. 

Motivated by an increasing interest in openness, transparency, and accountability in AI, the contributions of this work are: (i) improving the transparency of norms for collecting and sharing data related to accessibility and aging and (ii) calling the research accessibility community to action for guidelines and frameworks. Despite the critical role that data play to replicability, innovation, and a more inclusive future of AI, determining the answer to the questions, ``what data can I share and how?'' is yet not straightforward. 

\section{Related Work}
    We discuss and contrast our work with prior work on disability data and data sharing practices in related fields. 
    
    \subsection{Systematic Review of Disability Data}
       Prior efforts looking at cross-disability datasets have mainly focused on demographics, disability related diagnoses, causes of injury, interventions, outcomes, and costs (\eg BMS National Database~\cite{unk1994burn}). We also find systematic analyses of accessibility datasets and publications focusing in websites~\cite{thomsen2006building}, physical environments~\cite{ding2014survey}, and data collection and study methods~\cite{chamie1989survey, blaser2020why, markesich2021surveying, mack2021what}. Complimentary in nature to these efforts, this work focuses on a systematic review of datasets and related publications that source data from people with disabilities and older adults (\eg photos taken by blind users, video recordings of Deaf signers, and sensor data from gait movements of people with Parkinson's disease). What is important about these data is that they can be used in AI-infused applications and thus, require a delicate balancing act between benefits and risks for sharing. There are prior attempts reviewing such datasets. However, they tend to be more narrow in focus typically restricted to a population and task \eg sign language videos of Deaf signers~\cite{vaezijoze2019ms-asl, moryossef2021slp},  photos taken from people who are blind or low vision~\cite{lee2019hands}, and presence of older adults in publicly available face datasets~\cite{park2021understanding}.  
        
    \subsection{Data Sharing in Health and Accessibility}
    Datasets and data sharing have often served to attract, nurture, and challenge people in computing and technology. Many fields, including the health community, have seized this opportunity to promote AI in their area~\cite{diamond2009collecting, walport2011sharing, fecher2015what}.  We observe a similar trend in accessibility, with data challenges in the computer vision, speech, and broader machine learning communities (\eg Kaggle~\cite{kaggle}) designed around tasks and datasets from people with visual impairments (\eg~\cite{gurari2018vizwiz}) and Parkinson’s (\eg~\cite{schuller2015interspeech, kaggle2012smartphone}), respectively. 
    Even though accessibility and health data can share similarities (\eg smaller populations and sensitive information) risks may differ  (\eg accessing one's  mouse movements or clicks could be easier than having access to one's blood, saliva, or urea samples). More so, the health community has a longer history of discussions and frameworks around data ethics~\cite{xafis2019ethics}. For example, venues may require researchers to submit a data sharing plan when registering clinical trials and well-defined data sharing statements along their manuscripts~\cite{taichman2017data}. Such conversations around data artifacts are still ongoing in many computing communities (\eg~\cite{gebru2018datasheets, jo2020lessons, acm2020artifact}). The timing is right for the accessibility community to be involved in these discussions. 
    Within the broader human-computer interaction community, we see that prior work looking at 509 papers on wellness, accessibility, and aging published at ACM CHI 2010–2018, found that only 3 made their data publicly available~\cite{abbott2019local}. While not directly comparable, this number is quite low when compared to another work surveying CHI authors from the same period~\cite{wacharamanotham2020transparency}; researchers found that out of 373 reporting or generating any type of data, 80 shared raw data. Reasons for not sharing included data sensitivity, participant consent, and re-identification risks. We believe that this difference can be also explained by the increased privacy risks for accessibility data that are amplified by the risk of disability disclosure.

\section{Method}
To understand the current state of accessibility datasets and reflect on differences across communities of focus and disciplines involved in the data collection in the past, we perform a two-year long iterative process for dataset search, code, and review, which started in November 2017 and ended in March 2020. The scholars involved had varying levels of familiarity with accessibility and AI. This review includes a qualitative and descriptive analysis of the identified datasets spanning over 35 years (1984-2020, N=\datasetsTotal datasets). 

\subsection{Identifying  Accessibility Datasets}
Dataset search is a difficult task with some researchers arguing that it is a field on its own~\cite{chapman2020dataset}. Other factors making our search even more challenging are: datasets may not be publicly available; there is a vast terminology for communities of focus in accessibility and aging; and data collection tasks and purpose can highly vary.

\subsubsection{Criteria for Inclusion}
\label{section:conditions}
    Datasets had to satisfy the following:
    
    \noindent\textbf{Include data sourced from people with disabilities.} Many datasets may include people with disabilities without having explicit information about disability as about 15\% of the world's population live with some form of disability~\cite{WHO}. Given that there is no way for us to verify it, those datasets are not included in our collection. We include only datasets, where people with disability are explicitly mentioned to be contributing data. Terminology related to disability can be challenging. When in doubt, we consulted terminology in the IDEA Act~\cite{ideaact, idealist}, the ADA act~\cite{ada}, and accessibility research. We found that even when appropriate terms were explicitly mentioned they were ambiguous or misleading. Here are a few examples related to data from the Deaf community.  Sometimes, data stewards refer to those contributing data as ``participants'' and in other parts mention that they ``talked to the Deaf community'' without explicitly stating whether those contributing data were deaf/Deaf. The term ``signer'' is often used in the linguistic community to indicate native or fluent signers. However, there are many sign language datasets where the signers are not actually deaf/Deaf or hard-of-hearing people even though the term ``signer'' is used \eg to refer to interpreters~\cite{rwthfingerspelling}.  Also, it is not uncommon for those new to the field to collect data on their own or with other hearing people just by mimicking signs and still use the term ``signer'' without any other information~\cite{latif2019arasl, kaggleturkish}. Even when terms like ``fluent'' or ``expert'' are used, often the data collection process is missing information about the people contributing to the data thus preventing us from discovering if the experts or those fluent in a certain sign language were deaf/Deaf or hard of hearing versus sign language learners or hearing interpreters~\cite{bragg2019sign}. When we are not certain, datasets are not included. 

     \noindent\textbf{Include data that can be used in AI-infused applications}. We focus on datasets that include images, text, sensor data, and logs generated or reflective of people with disabilities. We focus on AI-infused applications and technologies for end users.
     Datasets that used sophisticated technology such as fMRI with people with dementia~\cite{mascali2015resting} or invasive technologies such as brain electrodes for enabling motor control, do not meet this criteria. While these technologies are promising, they are applicable to clinical settings. They would be out of reach for most of the general public and only be of interest to cliques within their respective research communities.

\subsubsection{Search Strategy} We employed a multilayer strategy for identifying relevant datasets: 
(i) \textit{open search} e.g., using specific keywords on search engines, (ii) \textit{focused searches in repositories and publication venues}, and (iii) \textit{focused searches on authors} that we found to have shared or provided contact information for sharing datasets in the past. This was not a linear process. We would go back and forth. 

For the \textbf{open search}, we used terms that are associated with the communities of focus such as \textit{disability}, \textit{disabled}, \textit{accessibility}, \textit{impairment}, \textit{visual}, \textit{disease}, \textit{speech}, \textit{hearing}, \textit{cognitive}, \textit{mobility}, \textit{vision}, \textit{autism}, \textit{behavorial}, \textit{developmental}, \textit{disorder}, \textit{learning}, \textit{cleft}, \textit{dyslexia}, dysarthia, Parkinson's disease, speech pathology, \textit{phonetic articulation}, \textit{sign language}, \textit{cleft lip}, \textit{Aspergers syndrome}, \textit{low vision}, \textit{depression}, and \textit{bipolar}. These were complemented with terminology used in the IDEA~\cite{ideaact, idealist} and The Americans with Disabilities~\cite{ada} Act along the with terms such as \textit{dataset}, \textit{database}, \textit{repository}, \textit{data}, \textit{collection}, \textit{eye tracking}, \textit{kinect}, and \textit{glove}.

For the \textbf{focused search in repositories and publication venues}, we started with known machine learning repositories such as Kaggle~\cite{kaggle}, UCI~\cite{bache2013uci}, and VisualData~\cite{feng2018visualdata}. We found that less than 0.15\% of available datasets in Kaggle were somehow related to disability and aging; the majority of them focused on healthcare rather than assistive or inclusive technologies. Only 4 of them met our criteria. Similar patterns were found in UCI and VisualData with 3 and 2 datasets meeting our criteria, respectfully. 
We also searched through digital libraries in scientific societies such as ACM~\cite{acm}, IEEE~\cite{ieee}, LREC~\cite{lrec}, ISCA~\cite{isca}, ACL~\cite{acl}, and CVF~\cite{cvf} as well as open science efforts such as Zenodo~\cite{zenodo}. Many of these venues do not typically provide a database or a search engine for the available datasets\footnote{ACM Digital Library offers a dataset search only for dataset artifacts; does not include datasets that authors may link to in their publications.}. When possible, we used the keywords from our open search or an HTML Regex addon, Find Plus~\cite{find2018richardson}, to find matches on lists of publications or datasets. As a last resort, we manually scanned publications, \eg, by zooming in on a specific venue such as ASSETS 2008-2018. In the midst of our research in 2018, we saw many efforts from the industry to help with the discovery and sharing of datasets such as the Registry of Open Data on AWS~\cite{awssearch}, Microsoft Research Open Data~\cite{microsoftsearch}, and the Google Search~\cite{googlesearch}, though, they resulted in very few new datasets in our collection. 

Last, for the \textbf{focused search on authors}, we used a seed list of authors resulting from the previous searches and started growing it organically.  
As we discovered more datasets, we identified common authors, principal investigators, and funded projects which we kept adding to our list of authors and keywords.

\subsection{Coding and Analyzing Accessibility Datasets}
We explore how data stewards across different disciplines and over time have balanced risks and benefits when collecting, reporting, and sharing accessibility data by looking at current and historical trends through an exploratory analysis. Deciding-what-to-explore-next is one of three key challenges in exploratory data analysis~\cite{lam2008framework}; in our analysis the question formulation were affected by (a) the feasible shape and structure of our codes, (b) the intended audience such as researchers and policy makers, and (c) our domain knowledge. For example, we wanted to explore data size in terms of number of data points per data contributor (a study participant in most datasets). However, accessibility datasets are so diverse in terms of datatype, granularity, and annotation, that having information on the number of data points per participant is not feasible and often not meaningful for a comparison. Even when looking at prior surveys on datasets within a community (\eg, sign language data), we see comparisons at a per participant level being avoided as the size of datasets could be measured either in sentences, individual signs, duration of continuous signing, or richness of linguistic annotations. 

\subsubsection{Manual Coding} We extract the following when available: \\
\textbf{About}: We note the name, year of release, links to the paper and/or dataset, DOI, data stewards (\eg, authors), and contact emails.\\
\textbf{Contributors}: We extract terms originally used to describe the individuals contributing data.\\
\textbf{Communities of Focus}: We assign a dataset to one of more communities of focus based on the data contributors (see Section~\ref{section:communities}).\\
\textbf{Collectors}: We note whether the dataset was collected by an educational institution, industry, or both. \\ 
\textbf{Access}: We scan the papers and/or online records to see whether there is a link for direct download or an explicitly mentioned point of contact for accessing the dataset as well as associated licensing.\\
\textbf{Ethical Board Clearance}: We mark whether ethical board clearance is mentioned in the datasets or their associated materials.\\
\textbf{Data Size}: We use the number of contributors as a proxy for data size. Control groups when available are coded separately.\\
\textbf{Data Type}: We make a note of the technology used to collect the data and list all formats both for the data and annotations if any.\\
\textbf{Summary}: We write a few sentence-long summary about the people, strategy, and purpose of the data collection.

\subsubsection{Communities of Focus}
\label{section:communities}
Rigid categorization of disability is difficult~\cite{blaser2020why, berg1992second} and perhaps a questionable task as conditions are vast and fluid~\cite{whittaker2019disability}. To aid the analysis and presentation, we annotated the datasets in our collection across 10 groups, with many communities falling under more than one. In the Results section, we augment discussions by analyzing the language used to describe the people within these communities (terms are indicated in quotes) and contrast them to current guidelines~\cite{hanson2015writing, guidelines2020research}. 

        \noindent\textbf{Autism.} This group includes datasets sourced from autistic individuals, both children and adults (\eg,~\cite{duan2019dataset} and ~\cite{eraslan2020autism}). The umbrella classification of \textit{Autism Spectrum Disorder (ASD)} refers to disorders of brain development affecting social interaction or verbal and nonverbal communication~\cite{grant2013proposed}. It is a group on its own because the  community has received attention as a specific subset of accessibility research~\cite{spiel2019agency, mack2021what} and in the IDEA Act~\cite{ideaautism}.
        
        \noindent\textbf{Cognitive.} This group captures a much broader category of datasets (\eg,~\cite{adams2017high} and ~\cite{garcia2018depresjon}) sourced from individuals with cognitive disorders commonly leading to different types of impairments such as Parkinson's disease, which can cause speech or mobility impairments, or Aphasia, which refers to a language impairment. We often cross-listed these datasets in multiple groups listed below. The group also considers other medical conditions that can lead to cognitive impairments such as bipolar disorder or functional memory deficit. 
        
        \noindent\textbf{Developmental.} This group broadly captures datasets from people with any physical and/or mental disability (\eg,~\cite{corrales2016use} and ~\cite{ferry2014diagnostically}) that began before the age of 22~\cite{hanson2015writing}. Though many people with a developmental disability do not have an intellectual disability~\cite{guidelines2020research}, the community is being noted as Intellectual or Developmental Disability (IDD)~\cite{mack2021what}. We group datasets that particularly mention and include individuals with IDD.  
        
        \noindent\textbf{Health.} This group refers to datasets sourced from people with a diverse demographic typically collected in research related to healthcare as well as health and wellbeing management (\eg,~\cite{leightley2015benchmarking} and ~\cite{dolatabadi2017toronto}),  though, we see intersections with other sub-disciplines of HCI, particularly aging and rehabilitation~\cite{abbott2019local, mack2021what}. Thus, this group includes older adults or people undergoing rehabilitation, as well as those with specific health concerns that do not belong to other groups such as cardiac diseases.
        
        \noindent\textbf{Hearing.} This group follows a combination of the IDEA Act category including datasets that involve people who are deaf/Deaf and hard of hearing. In this group, those who are deaf/Deaf often contribute as signers (\eg,~\cite{camgoz2016bosphorussign} and~\cite{huenerfauth2014release}).
        
        \noindent\textbf{Language.} This group includes datasets from people with language disorders such as aphasia or impairments represented by low verbal IQs that affect a person's ability to communicate. We group datasets that particularly focus on limited abilities to use and express language (\eg,~\cite{depaul2016corpus} and ~\cite{wetherell2007narrative}).
        
        \noindent\textbf{Learning.} This group includes datasets from people with  conditions that are neurologically-based and that can lead to difficulty in learning and using related skills~\cite{guidelines2020research}, such as reading or writing. We particularly group datasets that mention and include people with dyslexia, dysgraphia, or dyscalculia (\eg,~\cite{rello2015detecting} and ~\cite{lustig2016identifying}).
        
        \noindent\textbf{Mobility.} This group includes datasets from people with mobility and motor/dexterity impairments (\eg,~\cite{vatavu2019stroke} and ~\cite{anantharam2013predicting}). Although the word Mobility is used here for grouping, we generally include datasets that target people with limited physical functioning of one or more limbs, such as in walking or moving hands or fingers.
        
        \noindent\textbf{Speech.} This group covers datasets sourced from people with limited or impaired speech patterns, such as found in dysphonia, dysarthia, stuttering, and conditions of cleft lip and cleft palate. Since speech impairments can be caused by Parkinson's disease this group is cross-listed with other groups. Examples of data are ~\cite{vikram2019detection} and ~\cite{meunier2016typaloc}.
        
        \noindent\textbf{Vision.} This group includes dataset sourced from people who are blind or have low vision (\eg,~\cite{flores2018weallwalk} and ~\cite{gurari2018vizwiz}), which are commonly noted as the \textit{visually impaired} or those with \textit{vision loss}~\cite{mack2021what}.

\subsection{Reflections in Our Method and Limitations}
While we strived to be methodical in dataset search, it involved a lot of detective work. Thus, the dataset collection could be biased. The overseeing faculty, who has been working in the area of accessibility for more than 12 years, leveraged domain knowledge to complement the search by pointing to publication venues, data repositories, datasets, and data stewards. The initial search was performed by two Master's students in Information Management who were not familiar with accessibility and were guided through weekly meetings with the overseeing faculty. Then two HCI graduate students (Master's and PhD level) continued expanding the collection and started coding. They calculated inter rater agreement. Though this was found to be high, there were many errors in the coding. This is partially explained by the fact that the task was inherently messy and challenging. We see similar tasks, even when limited in scope within the field of accessibility and publication venues such as ACM, characterized as ``challenging and effortful''~\cite{mack2021what}.

We found disability-related terminology to be a dual challenge. First, it was difficult to come up with an extensive list of past and current terms than can be used as keywords for finding datasets as data stewards may have used depreciated terminology or more fine-grained terms established within a field to refer to specific sub populations. Second, it was difficult to map terms used in datasets and/or associated publications to the communities of focus. Partially, this is explained by the fact that in the broader research community, terminology used to describe the people contributing data is often confusing, as discussed above. However, we believe that the lack of extensive accessibility experience may have also contributed. The team would meet often to discuss challenging cases and resolve disagreements. A fifth student, pursuing a doctoral program,  was added to the team; the student had a few years of experience in accessibility. The student had a detailed pass on the annotations, resolved any conflicts with the first PhD student, confirmed difficult cases with the faculty, and helped co-lead the analysis efforts. 

\section{Results}
We characterize the current status of accessibility datasets (1984-2020) in terms of the communities of focus, current distribution of the datasets across communities and how that changed over the years, data collection purpose, and language used to describe the people who contributed data. We report trends across data size in terms of the number of people involved referring to both those representing communities of focus and those serving as proxies or control~\cite{mack2021what}. We explore how data types relate to communities and the purpose of the data collection. Last, we identify common data sharing practices and report on mentions of research clearance.

\subsection{Communities of Focus}
    \begin{figure}
    \includegraphics[width=1\linewidth]{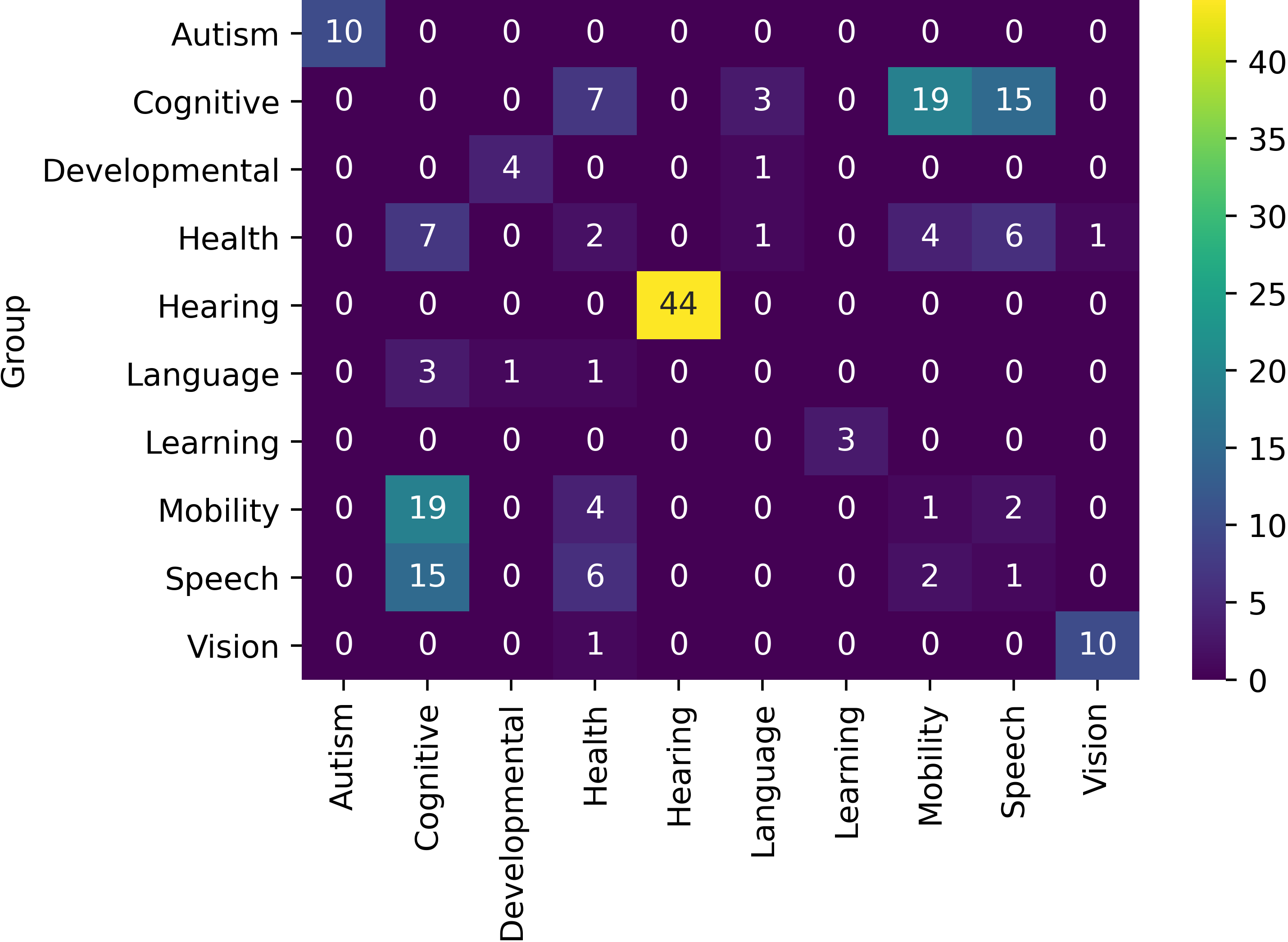}
    \vspace*{-5mm}
    \caption{Distribution of dataset count across all communities.}
    \label{fig:communities}
    \Description[Distribution of datasets over communities, columns are ordered as follows, each row begins with the name of the group followed by the number of dataset for each group it intersects with, since a dataset can belong to multiple communities.]{
                   Autism  Cognitive  Developmental  Health  Hearing  Language  \
Group                                                                        
Autism         10      0          0              0       0        0          
Cognitive      0       0          0              7       0        3          
Developmental  0       0          4              0       0        1          
Health         0       7          0              2       0        1          
Hearing        0       0          0              0       44       0          
Language       0       3          1              1       0        0          
Learning       0       0          0              0       0        0          
Mobility       0       19         0              4       0        0          
Speech         0       15         0              6       0        0          
Vision         0       0          0              1       0        0          

               Learning  Mobility  Speech  Vision  
Group                                              
Autism         0         0         0       0       
Cognitive      0         19        15      0       
Developmental  0         0         0       0       
Health         0         4         6       1       
Hearing        0         0         0       0       
Language       0         0         0       0       
Learning       3         0         0       0       
Mobility       0         1         2       0       
Speech         0         2         1       0       
Vision         0         0         0       10      
    }
    \end{figure}

\begin{figure}

    \includegraphics[width=0.9\linewidth]{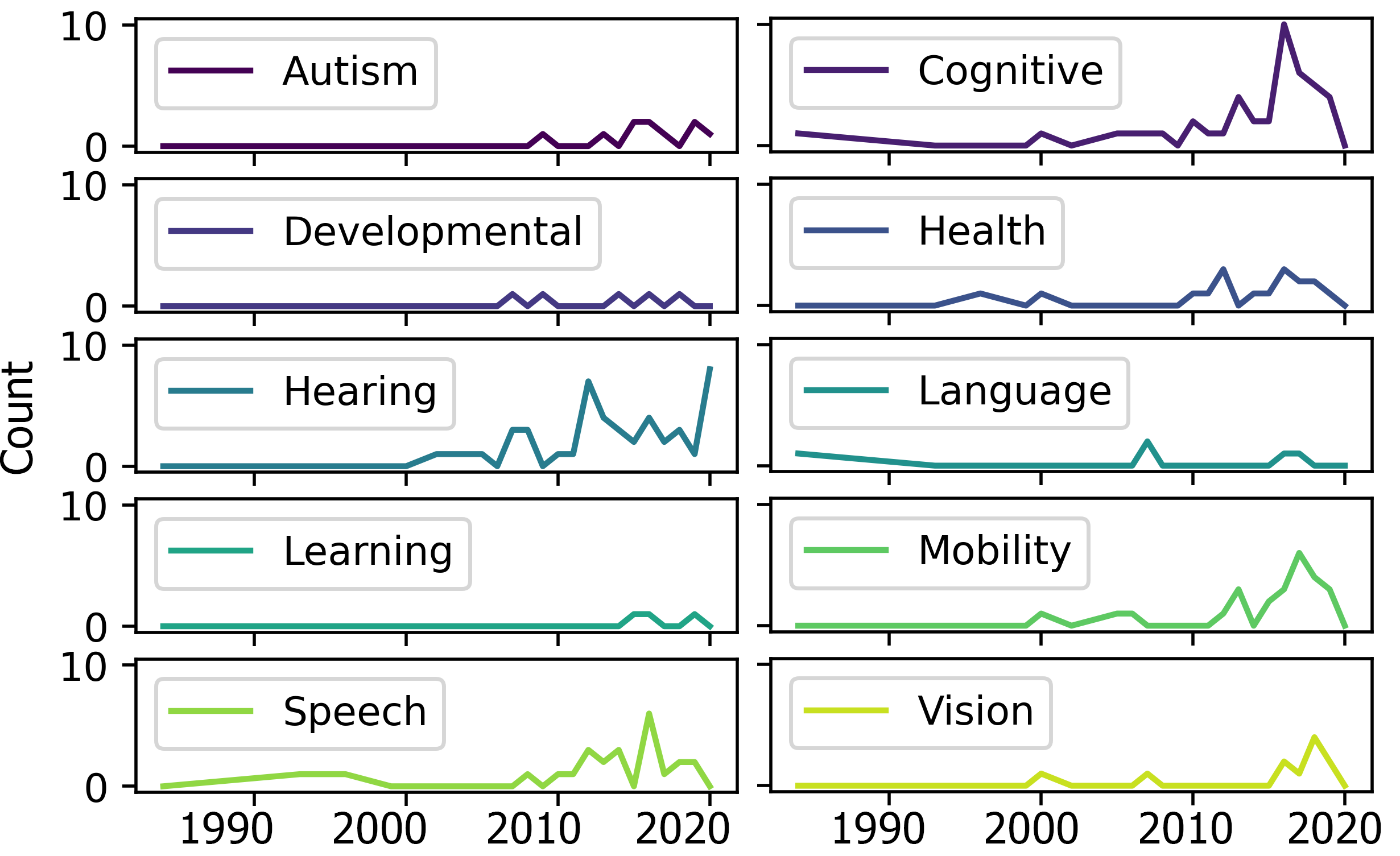}
    \caption{Dataset count over the years across all communities.}
    \Description[List of the number of datasets every year for each group, from the year 1984 to 2020]{
       Autism  Cognitive  Developmental  Health  Hearing  Language  Learning  \
Year                                                                          
1984  0       1          0              0       0        1         0          
1993  0       0          0              0       0        0         0          
1996  0       0          0              1       0        0         0          
1999  0       0          0              0       0        0         0          
2000  0       1          0              1       0        0         0          
2002  0       0          0              0       1        0         0          
2005  0       1          0              0       1        0         0          
2006  0       1          0              0       0        0         0          
2007  0       1          1              0       3        2         0          
2008  0       1          0              0       3        0         0          
2009  1       0          1              0       0        0         0          
2010  0       2          0              1       1        0         0          
2011  0       1          0              1       1        0         0          
2012  0       1          0              3       7        0         0          
2013  1       4          0              0       4        0         0          
2014  0       2          1              1       3        0         0          
2015  2       2          0              1       2        0         1          
2016  2       10         1              3       4        1         1          
2017  1       6          0              2       2        1         0          
2018  0       5          1              2       3        0         0          
2019  2       4          0              1       1        0         1          
2020  1       0          0              0       8        0         0          

      Mobility  Speech  Vision  
Year                            
1984  0         0       0       
1993  0         1       0       
1996  0         1       0       
1999  0         0       0       
2000  1         0       1       
2002  0         0       0       
2005  1         0       0       
2006  1         0       0       
2007  0         0       1       
2008  0         1       0       
2009  0         0       0       
2010  0         1       0       
2011  0         1       0       
2012  1         3       0       
2013  3         2       0       
2014  0         3       0       
2015  2         0       0       
2016  3         6       2       
2017  6         1       1       
2018  4         2       4       
2019  3         2       2       
2020  0         0       0    }
    ~\label{fig:communityOverTime}
    \end{figure}

In Figure~\ref{fig:communities}, we see that at a high level the representation across communities is not equally distributed with datasets sourced from the Deaf community and broader Cognitive group dominating; followed by datasets related to communities within the Mobility and Speech groups, which are often cross listed with Cognitive. Surprisingly, the disproportionate attention that communities within the Vision group have received in accessibility research~\cite{mack2021what} is not reflected in datasets, though we see an uptick in more recent years (Figure~\ref{fig:communityOverTime}).

\subsubsection{Autism}
\newcommand\autTotal{10} 
\newcommand\autChild{5} 
\newcommand\autAdult{5} 
\newcommand\autHighFunction{5} 
\newcommand\autAutistic{2} 
\newcommand\autWithAutism{7} 

Datasets under this group (a total of \autTotal) starting around 2010, include data both from children (\autChild) and adults (\autAdult). 
Data from children focus mostly on early detection through eye-tracking measurements~\cite{carette2019learning}, vocalization composition~\cite{xu2009automatic}, or eye-contact detection in dyadic interactions~\cite{rehg2014behavioral}. Only one relates to analyzing behaviors of participants interacting with technology such as robots~\cite{leo2015automatic}. Data from adults focus on increasing accessibility, such as browsing~\cite{eraslan2020autism} and text readability~\cite{yaneva2015accessible}. When describing the people contributing data, we find  that terms such as ``\textit{person with autism}''  or ``\textit{person with ASD}'' were the most common (\autWithAutism), in contrast to recommended language for this community, either identity-first (\eg, autistic person) or person-first (\eg, person on the autism spectrum)~\cite{guidelines2020research}. Datasets often resulted from studies assessing differences across autistic and non-autistic participants. Non-autistic individuals were described as ``\textit{neurotypical}'' or ``\textit{non-ASD/non-autistic}'' by accessibility and health informatics researchers~\cite{eraslan2020autism,carette2019learning} and as ``\textit{healthy controls}`` or ``\textit{typically developing}'' by others~\cite{duan2019dataset, xu2009automatic}. 
There were few occurrences of misnomers
such as ``\textit{high functioning}'' which open up the risk of inaccurate labels to describe autistic individuals~\cite{alvares2020misnomer}. Surprisingly, they were present both in broader computing and accessibility publications~\cite{duan2019dataset, eraslan2019web}.

\subsubsection{Cognitive}

\newcommand\cogTotal{44} 
\newcommand\cogNames{42} 
\newcommand\cogMobility{19} 
\newcommand\cogSpeech{15} 
\newcommand\cogPD{29} 
\newcommand\cogDementia{6} 
\newcommand\cogPatients{31} 

Datasets under this group (a total of \cogTotal) also start around 2010 with a few occurrences as early as 1985. They encompass a wide range of communities typically including people with Parkinson's (PD)~\cite{orozco2016automatic}, Dementia~\cite{makimoto2006japankorea}, Dysarthria~\cite{fougeron2010developing}, Alzheimer's~\cite{kempler1987syntactic}, Aphasia~\cite{macwhinney2011aphasiabank}, Huntington's (HD)~\cite{novotny2016hypernasality}, Amyotrophic Lateral Sclerosis (ALS)~\cite{hausdorff2000dynamic}, Episodic Memory Impairment~\cite{lee2007providing}, Ataxia~\cite{jaroensri2017video}, or Bipolar disorder~\cite{garcia2018depresjon}. While the majority of datasets focus on Parkinson's disease (\cogPD) we often see a combination (\eg, PD and HD~\cite{novotny2016hypernasality}; ALS, PD and HD~\cite{hausdorff2000dynamic}) with a shared goal of identifying ``\textit{neurodegenerative}'' signs underlying these conditions, commonly in speech or mobility. 
This is why in Figure~\ref{fig:communities} we often cross list datasets in this group with Mobility (\cogMobility) and/or Speech (\cogSpeech). Guidelines recommend for cognitive disabilities to be defined clearly~\cite{hanson2015writing}. We see this being a common trend (\cogNames), except when referring to ``\textit{people with mild cognitive impairment,}'' a diagnostic category in which its formal definition and measures are the field of inquiry~\cite{quinn2020terms}. Given that communities in this group have been actively studied in health and clinical domains for a long time, we see more organized data sharing efforts such as TalkBank~\cite{macwhinney2004talkbank} released in early 2000s followed by AphasiaBank~\cite{macwhinney2011aphasiabank}.

\subsubsection{Developmental} 
\newcommand\devTotal{5} 
\newcommand\devIntellectual{4} 
\newcommand\devChild{3} 

Datasets under this group (a total of \devTotal) shyly start appearing before 2010 and include data from people with a specific diagnosis related to developmental disabilities (\eg, Down syndrome)~\cite{ferry2014diagnostically, corrales2016use} and those without a known diagnosis whose disabilities are described as moderate or mild intellectual disabilities~\cite{feng2009automatic, corrales2016use, aggarwal2018evaluation}. The majority include data from children (\devChild). Developmental disability represents a broader category of often lifelong disability that can be intellectual and/or physical~\cite{hanson2015writing}. The term ``\textit{intellectual disabilities},'' often noted as (ID), was found in studies that referred to participants' limitations in cognitive functioning and adaptive skills (\devIntellectual), such as communication~\cite{aggarwal2018evaluation} or reading~\cite{feng2009automatic}.  The datasets associated with studies in ID are typically used in machine learning models to find diagnostic predictors~\cite{aggarwal2018evaluation, feng2009automatic}. The term ``\textit{developmental disorders}'' was found in a study focusing on the identification of what authors call ``ultra-rare developmental diseases'' using facial phenotypes from photographs~\cite{ferry2014diagnostically}. One of the datasets includes adolescents with language impairments and low non-verbal IQ~\cite{wetherell2007narrative}; this dataset is thus cross-listed in Figure~\ref{fig:communities} with Language and Developmental. In contrast to other groups, we do not see a growth over the years and intent of data sharing (see Section 4.4). 



\subsubsection{Health}
\newcommand\healTotal{17} 
\newcommand\healCog{7} 
\newcommand\healSpeech{6} 
\newcommand\healMobility{4} 
\newcommand\healPD{5} 
\newcommand\healOnly{3 }

Datasets under this group (a total of \healTotal) with peaks in the 90s and this past decade, include data from people undergoing stroke rehabilitation~\cite{dolatabadi2017toronto}, people with depression~\cite{garcia2018depresjon}, retinopathy disorders~\cite{hoover2000locating}, levodopa-induced dyskinesia~\cite{li2018vision}, dysarthric speech after neck cancer surgery~\cite{clapham2012nki}, or suspected dementia~\cite{becker1994natural}. Datasets are often cross-listed with Cognitive (\healCog), Speech (\healSpeech), and Mobility (\healMobility). The overall space of Health is vast, but within the scope of HCI, studies have been centered around the impact of technologies on the practices and experiences of health professionals and patients~\cite{blandford2019hci}. Similarly, datasets in this group relate to applications in healthcare settings, incorporating telehealth mechanisms or automatic patient screening (\eg, predicting Parkinson's disease progression with smartphone data~\cite{anantharam2013predicting}). Thus, their increase in the last decade could be explained by these research directions. In contrast, in the 90s the emphasis was on clinical studies analyzing dysarthric speech~\cite{menendez1996nemours} or retinal images~\cite{hoover2000locating}. Datasets from research in aging are also prevalent in this group, where terms ``\textit{older adults}'' and the ``\textit{elderly},'' which the latter is rather deprecated in the community~\cite{lundebjerg2017comes}, are used. The purpose of the data collection often relates to understanding age-related factors, such as motor movements~\cite{leightley2015benchmarking,moffatt2010addressing}. Although these datasets represent a highly diverse demographic in terms of aging~\cite{knowles2019hci}, we did not observe a consistent age being used as a threshold.







\subsubsection{Hearing}
\newcommand\hearTotal{44} 
\newcommand\hearASL{12}
\newcommand\hearGerman{8}
\newcommand\hearFrench{6}

Datasets under this group (a total of \hearTotal) starting in year 2000, include data from people who are deaf/Deaf or hard of hearing, typically consisting of sign language videos and gloss annotations. Even though this is the most represented group in terms of the number of datasets, researchers in related fields still call for more datasets: larger, more representative, and public~\cite{bragg2019sign}. This group is also the most diverse in terms of the research communities involved in the data collection process including computer vision~\cite{bull2020mediapi}, linguistics and natural language processing~\cite{neidle2012challenges, ebling2018smile}, as well as accessibility and human-computer interaction~\cite{huenerfauth2014release}. However, as discussed in Bragg \etal~\cite{bragg2019sign}, data collection still occurs in separate disciplinary silos. We find that  
the most common sign languages elicited were American (\hearASL), German (\hearGerman), and French (\hearFrench); the rest were Polish, Greek, Chinese, Finnish, British, Bangla, Turkish, Czech, Auslan (Australian sign language), Libras (Brazilian sign language), Arabic, Flemish, Spanish, Korean, Russian, and Swiss-German Sign Language. Very few datasets targeted multiple sign languages~\cite{cooper2012sign, matthes2012dicta,toman2016complexity} to establish machine learning benchmarks.  When referring to signers contributing data, terms like ``\textit{Native}'', ``\textit{Expert}'', and ``\textit{Fluent}'' were often used.

\subsubsection{Language}
\newcommand\langTotal{5} 
\newcommand\sizeLangTotal{5 } 
\newcommand\langAphasia{3}
\newcommand\langAdults{4 }

Datasets under this group (a total of \langTotal) appearing early in 1985 and then much later in 2005, include people with aphasia~\cite{kempler1987syntactic, depaul2016corpus} or language impairments who have intact or lowered non verbal IQ~\cite{wetherell2007narrative}. Aphasia is a disorder of linguistic processing related to specific brain regions~\cite{damasio1992aphasia}. Hence, datasets related to Aphasia are cross-listed with Cognitive, though,  not always restricted to people exclusively with aphasia. All datasets in this group involve adults~\cite{sebastian2018patterns,depaul2016corpus} except one, which focuses on understanding the impact of language impairments on narrative skills of children as they reach adolescence~\cite{wetherell2007narrative}. Most, collected by researchers in Speech and Hearing Science or Aphasiology, aimed at identifying patterns of linguistic decline~\cite{macwhinney2011aphasiabank,kempler1987syntactic,sebastian2018patterns} with the intent to share data (\eg through AphasiaBank) only for specific research purposes. In contrast, datasets collected among accessibility researchers for purposes of developing augmentative and alternative communication devices~\cite{allen2007design} did not indicate any intent for data sharing.



\subsubsection{Learning} 
\newcommand\learnTotal{3} 

Datasets under this group (a total of \learnTotal) started appearing in 2015 and include data from children and adults with dyslexia;  with the target population being often children from 7 to 17 years old~\cite{rello2019predicting}. This is perhaps not a surprise as dyslexia is the most common neurobehavioral disorder related to children's learning ability~\cite{shaywitz1998dyslexia}. Datasets typically aim to support screening of dyslexia~\cite{rello2015detecting} and sometimes involved eye tracking to build predictive models~\cite{lustig2016identifying, rello2015detecting}. Data collection approaches vary significantly from gamified exercises eliciting data from a large pool of dyslexia-diagnosed children~\cite{rello2019predicting} to studies involving university students despite the work's motivation on children with dyslexia~\cite{lustig2016identifying}. 

\subsubsection{Mobility}
\newcommand\mobilityTotal{27} 
\newcommand\mobilityPD{21}

Datasets under this group (a total of \mobilityTotal) appearing around year 2000 and quickly growing, include hand or gait movement from people with motor impairments, mostly (\mobilityPD) found in studies related to Parkinson's disease~\cite{vasquez2018multimodal,white2019population} or dyskinesia~\cite{li2018vision}. Dyskinesia is commonly seen in Parkinson's disease patients after prolonged treatment with levodopa, often noted as levodopa-induced dyskinesia (LID), causing involuntary and uncontrollable movements~\cite{bezard2001pathophysiology}. Often studies include people with not only PD or LID but also various motor impairments, such as cerebral palsy~\cite{vatavu2019stroke}, spinal cord injury~\cite{mamem}, or spinal muscular atrophy~\cite{vatavu2019stroke}. Datasets vary in the sensing modalities, ranging from mouse cursor movements~\cite{white2019population}, wearable sensors~\cite{thomas2018treatment}, or stroke-gesture input on touchscreens~\cite{vatavu2019stroke}, to stride measurements from insole sensitive resistors~\cite{hausdorff2000dynamic,vasquez2018multimodal} or vision-based pose estimation~\cite{li2018vision}. We observe that over the last decade datasets in this group are often shared.

\subsubsection{Speech}
\newcommand\speechTotal{24}
\newcommand\sizeSpeechTotal{24 } 
\newcommand\sizeSpeechOnly{7 } 
\newcommand\speechPD{12}
\newcommand\speechDysarthria{11}

Datasets under this group (a total of \speechTotal) appearing since the 90s, include mainly people with dysarthia (\speechDysarthria) and Parkinson's disease (\speechPD) with some overlap between the two. The other datasets included samples of impaired speech due to neck cancer~\cite{clapham2012nki}, cleft lip~\cite{vikram2019detection}, and other causes of pathological speech like Amyotrophic Lateral Sclerosis (ALS)~\cite{rudzicz2012torgo}. Given that communities in this group have been actively studied for a long time, we see more organized data sharing efforts, common in the speech and language processing research community, such as creating dysarthric speech databases for speech recognition. Dysarthria is a motor speech disorder due to neurological disease or injury, and those with dysarthria experience difficulty in articulating words~\cite{darley1969differential}. Many have build benchmark datasets for dysarthria since early 90s~\cite{deller1993whitaker, menendez1996nemours, kim2008dysarthric} with others following in more recent years~\cite{choi2012dysarthric, rudzicz2012torgo}. Dysarthric speakers contributing their data often have cerebral palsy~\cite{deller1993whitaker} and some have Parkinson's disease~\cite{fougeron2010developing} or ALS~\cite{rudzicz2012torgo}. However, the speech and language community still calls for more datasets with more appropriate sizes for machine learning tasks~\cite{fraser2019importance}.

\subsubsection{Vision}
\newcommand\visionTotal{11}
\newcommand\visionBlind{6}

Datasets under this group (a total of \visionTotal) typically include people who are blind~\cite{ahmetovic2018turn, bigham2007webinsitu} (\visionBlind) or have low vision~\cite{vatavu2018impact}. When datasets are collected with real-world assistive applications, where disability status as well as visual acuity or age of onset are not known, the umbrella term, ``\textit{people with visual impairments}'' or ``\textit{visually impaired}'' is used to describe those contributing data (\eg~\cite{kacorri2016supporting}). Datasets are typically collected in the context of accessibility such as navigation~\cite{flores2018weallwalk}, object recognition~\cite{sosa2017hands}, and accessibility of web or touchscreen interfaces~\cite{bigham2007webinsitu, vatavu2018impact}. There is one exception, where the context is clinical, focusing on screening of Proliferative Diabetic Retinopathy based on retina images~\cite{hoover2000locating}.  Although datasets in this group involve communities dominating accessibility research~\cite{mack2021what}, they are typically not shared.  This might seem surprising. However, we suspect that it merely mirrors local data sharing standards in the accessibility community, where dataset contributions are rare~\cite{mack2021what, abbott2019local}. Perhaps, this could be explained by the awareness that many accessibility researchers have towards potential risks but also the lack of guidelines and frameworks for ethical data sharing~\cite{morris2020ai}. 
Surprisingly, we did not find any datasets sourced from deaf-blind people, even though in the United States only, this population is estimated to be as high as 2.4 million~\cite{hellenkeller}.





\begin{figure*}[t]
    \includegraphics[width=1\linewidth]{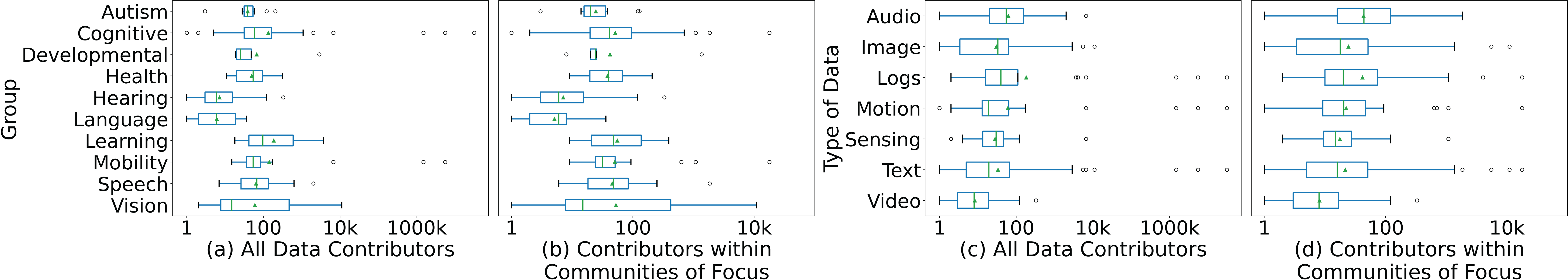}
    
    \centering
    \Description[Description of the boxplots for the number of participants who provided the data for each group. First its the data for (a) shows the total of participants who contributed and then (b) shows the contributions from participants within the communities of focus]{
    For figure (a) All data contributors split by Groups
number of people        Vision       Speech      Mobility     Learning  \
count             11.000000     24.000000    2.500000e+01  3.000000      
mean              1889.272727   178.041667   2.810337e+05  1253.000000   
std               3603.274902   410.510655   1.132748e+06  2071.043457   
min               2.000000      7.000000     1.500000e+01  18.000000     
25\%               8.000000      26.250000    3.600000e+01  57.500000     
50\%               15.000000     67.500000    5.400000e+01  97.000000     
75\%               2054.500000   134.750000   8.400000e+01  1870.500000   
max               11045.000000  2010.000000  5.526868e+06  3644.000000   

number of people   Language     Hearing      Health  Developmental  \
count             5.000000   43.000000   15.000000   5.000000        
mean              12.800000  21.860465   80.333333   598.000000      
std               14.822281  53.256603   86.462764   1274.613079     
min               1.000000   1.000000    11.000000   19.000000       
25\%               2.000000   3.000000    20.000000   20.000000       
50\%               6.000000   6.000000    54.000000   25.000000       
75\%               19.000000  15.500000   104.000000  48.000000       
max               36.000000  330.000000  312.000000  2878.000000     

number of people     Cognitive      Autism  
count             4.400000e+01  10.000000   
mean              8.716570e+05  60.000000   
std               4.774916e+06  59.423153   
min               1.000000e+00  3.000000    
25\%               3.150000e+01  31.500000   
50\%               5.950000e+01  39.000000   
75\%               1.615000e+02  54.250000   
max               3.132177e+07  205.000000  

for Figure (b) Contributors within Communities of Focus split by Groups
number of people        Vision      Speech      Mobility    Learning  \
count             11.000000     24.00000    25.000000     3.000000     
mean              1886.727273   135.00000   814.160000    149.666667   
std               3604.711336   369.78325   3555.639859   210.770808   
min               1.000000      6.00000     9.000000      9.000000     
25\%               8.000000      18.25000    24.000000     28.500000    
50\%               15.000000     48.00000    32.000000     48.000000    
75\%               2049.500000   84.50000    51.000000     220.000000   
max               11045.000000  1850.00000  17843.000000  392.000000   

number of people   Language     Hearing      Health  Developmental  \
count             5.000000   43.000000   16.000000   5.000000        
mean              10.600000  21.790698   59.250000   288.000000      
std               14.484474  53.270450   58.570186   600.981281      
min               1.000000   1.000000    9.000000    8.000000        
25\%               2.000000   3.000000    19.750000   20.000000       
50\%               6.000000   6.000000    40.000000   24.000000       
75\%               8.000000   15.500000   68.750000   25.000000       
max               36.000000  330.000000  208.000000  1363.000000     

number of people     Cognitive      Autism  
count             44.000000     10.000000   
mean              563.477273    40.700000   
std               2687.749387   45.406926   
min               1.000000      3.000000    
25\%               19.750000     15.750000   
50\%               41.000000     20.000000   
75\%               94.250000     35.750000   
max               17843.000000  129.000000  

Figures c and d are for type of data, 
For figure c all data contributors split by type of data
number of people       Video          Text      Sensing        Motion  \
count             48.000000   9.200000e+01  23.000000    3.300000e+01   
mean              24.187500   4.170940e+05  323.347826   1.162040e+06   
std               52.383876   3.311349e+06  1383.431196  5.503583e+06   
min               1.000000    1.000000e+00  2.000000     1.000000e+00   
25\%               3.000000    5.000000e+00  13.500000    1.300000e+01   
50\%               8.000000    1.950000e+01  30.000000    1.900000e+01   
75\%               19.250000   6.725000e+01  47.000000    6.400000e+01   
max               330.000000  3.132177e+07  6668.000000  3.132177e+07   

number of people          Logs         Image        Audio  
count             2.500000e+01  22.000000     39.000000    
mean              1.534187e+06  911.181818    330.076923   
std               6.307996e+06  2606.825320   1098.345919  
min               2.000000e+00  1.000000      1.000000     
25\%               1.600000e+01  3.500000      20.000000    
50\%               4.000000e+01  34.000000     55.000000    
75\%               1.110000e+02  62.750000     154.500000   
max               3.132177e+07  11045.000000  6668.000000  

and Figure d shows contributors within the communities of focus split by type of data
number of people       Video          Text     Sensing        Motion  \
count             48.000000   92.000000     23.00000    34.000000      
mean              23.687500   472.760870    69.26087    616.823529     
std               52.423462   2244.982045   223.32207   3052.795713    
min               1.000000    1.000000      2.00000     1.000000       
25\%               3.000000    5.000000      9.50000     9.250000       
50\%               8.000000    16.000000     15.00000    20.500000      
75\%               17.000000   51.000000     27.50000    47.000000      
max               330.000000  17843.000000  1087.00000  17843.000000   

number of people          Logs         Image        Audio  
count             25.000000     22.000000     39.000000    
mean              1003.160000   836.636364    139.230769   
std               3604.964779   2574.017234   333.071864   
min               2.000000      1.000000      1.000000     
25\%               10.000000     3.500000      16.000000    
50\%               20.000000     18.000000     44.000000    
75\%               74.000000     51.500000     120.500000   
max               17843.000000  11045.000000  1850.000000  

}
    \caption{Sample size across communities and data types: all contributors (a\&c) vs. those within the communities of focus only (b\&d).}~\label{fig:boxplotDisability}
\end{figure*}
\subsection{Sample Size}
\newcommand\sizeSmallest{1 }
\newcommand\sizeLargest{17,843 }
\newcommand\sizeNA{2}

\newcommand\sizeControl{58 } 
\newcommand\sizeControlMin{1 }
\newcommand\sizeControlMax{31,321,070 }

Figures~\ref{fig:boxplotDisability}a and \ref{fig:boxplotDisability}b show the number of people (\textit{N}) contributing data across groups; Figure~\ref{fig:boxplotDisability}a includes those who served as a control and Figure~\ref{fig:boxplotDisability}b does not. Figures~\ref{fig:boxplotDisability}c and \ref{fig:boxplotDisability}d also show the sample size with and without the control respectively, this time across data types. Distributions are visualized as boxplots with a 1.5 interquartile range (IQR) on a logarithmic scale. For comparison, means are denoted with a triangle. We see that \textit{N} is highly variable across and within groups and that those serving as control are contributing to outliers (above 1000K).  Looking at Figure~\ref{fig:boxplotDisability}b, where data are sourced only from those within the communities of focus, we find that overall reported size ranged from \sizeSmallest to \sizeLargest  (median=20, IQR=51-8). Only few (\sizeNA) datasets do not provide information on the number of data contributors. Within the  \sizeControl datasets that included control groups (\eg, participants with and without dysarthric speech~\cite{meunier2016typaloc}), the reported size ranged from \sizeControlMin to \sizeControlMax (median=31, IQR=68-34). 

\newcommand\sizeLessHundred{115}
\newcommand\sizeOverThousand{7}
\newcommand\sizeMedian{20 }
\newcommand\sizeOne{7 }
\newcommand\sizeTwo{6 }
\newcommand\sizeThree{8 }
\newcommand\sizeOneTwoThree{21}

\newcommand\sizeHearingOneTwoThree{15}
\newcommand\sizeHearingLargest{330 }
\newcommand\sizeHearingMedian{6 }

\newcommand\sizeLanguageMedian{6 }

\newcommand\sizeSpeechMedian{48 }
\newcommand\sizeSpeechSmallest{6}
\newcommand\sizeSpeechLargest{1,850 }
\newcommand\speechControl{17 } 
\newcommand\sizeSpeechControlMedian{40 }

\newcommand\sizeCognitive{}
\newcommand\sizeMobility{}
\newcommand\sizeHearing{}
\newcommand\sizeSpeech{}

In this section, we focus our analysis on sample sizes that refer only to the number of participants within the communities of focus (Figure~\ref{fig:boxplotDisability}b).  We observe \textit{N}=3 being the most common sample size (\sizeThree  datasets);   \textit{N}=1 and \textit{N}=10 were second most common (each across \sizeOne datasets); and \textit{N}=2 being next in line (\sizeTwo datasets). The majority of datasets (\sizeLessHundred) have \textit{N}$\leq$100 and \textit{N}$\geq$1000 for only a few (\sizeOverThousand). 

Across all the communities of focus, the Speech group had the largest median value of \sizeSpeechMedian (85-18). Often studies in this group involved clinical trials for data collection that were performed through collaboration with the local hospital and medical lab, and/or the federation associated with the target disability group. In the extensive project by ~\cite{fougeron2010developing}, in partnership with multiple institutions including a hospital and a team of doctors for over 30 years, they produced two corpora having the largest population size of dysarthric speech data, with a total of \sizeSpeechLargest participants diagnosed with various neurological disorders. In comparison, Whitaker database of dysarthric speech due to cerebral palsy~\cite{deller1993whitaker} had the smallest sample size (\sizeSpeechSmallest) and was created by two research teams from different universities.

Although the Language group also collected speech samples like the Speech group, their median sample size was one of the smallest among our communities of focus. The median was \sizeLanguageMedian(8-2), the same value as the Hearing group of \sizeHearingMedian(16-3). This could be due to a lack of clinical collaboration. The largest sample size in the Language group was 36 in ~\cite{sebastian2018patterns}, being the only dataset (out of \sizeLangTotal) that was created in collaboration with a medical institution specialized in the target population (\ie, Aphasia). It also could be due to the complexity of speech tasks associated with the speech recordings. Studies in the Speech group often include a series of short tasks to collect vowel or phonation sounds following a limited set of acoustic measures (\eg, ~\cite{fougeron2010developing}, ~\cite{cesari2018new}), whereas studies in the Language group include story telling and conversational tasks to asses verbal and non-verbal communication skills that are more complex to analyze (\eg, ~\cite{sebastian2018patterns}, ~\cite{wetherell2007narrative}). The population pool size could also impact the availability of participants; for example, ~\cite{allen2007design} in the Language group collected data from aphasic participants but ~\cite{fougeron2010developing} in the Speech group collected dysarthric speech data from participants with various neurological diagnoses, such as Parkinson's, Amyotrophic Lateral Sclerosis, and cerebellar diseases. 

\subsubsection{N $>$ 1000}
When looking at datasets sourced from more that 1000 participants (a total of \sizeOverThousand), we observe that typically a remote data collection in the real world is involved, such as tracking device usage or asking people to upload data regularly using an app. 
Examples include the mPower study~\cite{bot2016mpower} that involved 1087 Parkinson's disease diagnosed individuals to record and register daily 4 activities via a mobile application: tapping, memory test, walking and voice samples of vowel pronunciations.
Similarly, the iMove and Vizwiz datasets were collected with assistive applications deployed in the real-world with 4055 and 11045 participants, respectively~\cite{kacorri2016supporting, bigham2010vizwiz}.
The largest sample size was 17,843, where mouse and keyboard interactions with a search engine were collected and users were classified as having Parkinson's disease based on proxy information from web search queries (\eg, ``I have Parkinson’s'')~\cite{youngmann2019machine}.

\subsubsection{N $\leq$ 3}
Many datasets (a total of \sizeOneTwoThree) had a small sample size, with data generated from no more than 3 people. The majority of them (\sizeHearingOneTwoThree) involved sign language recordings (\eg~\cite{kadous2002temporal, warchol2019recognition}). Unlike the large datasets above, where membership to a community of focus was often inferred on proxy information (\eg, use of a screen reader to access an assistive technology app or relevant search terms), these small datasets typically included rich data from series of tasks in controlled settings, where researchers had detailed information about those contributing to the data. For example, for sign language datasets this typically involves lab recordings in a fixed multi-camera setup with additional apparatus used to capture and process data. Given the difficult and costly process of these data collection methods, where people are explicitly asked to complete a series of tasks, it is not surprising to see the same people contributing to studies performed by the same research team (\eg,~\cite{oszust2013polish} and~\cite{kapuscinski2015recognition}). Depending on the amount of linguistic analysis required, such as gloss and non-manual annotations, these datasets could be constrained by the smaller pool of individuals who beyond sign language fluency also have computational linguistic training.

\subsection{Data Types}

\begin{figure}[t]
    \includegraphics[width=1\linewidth]{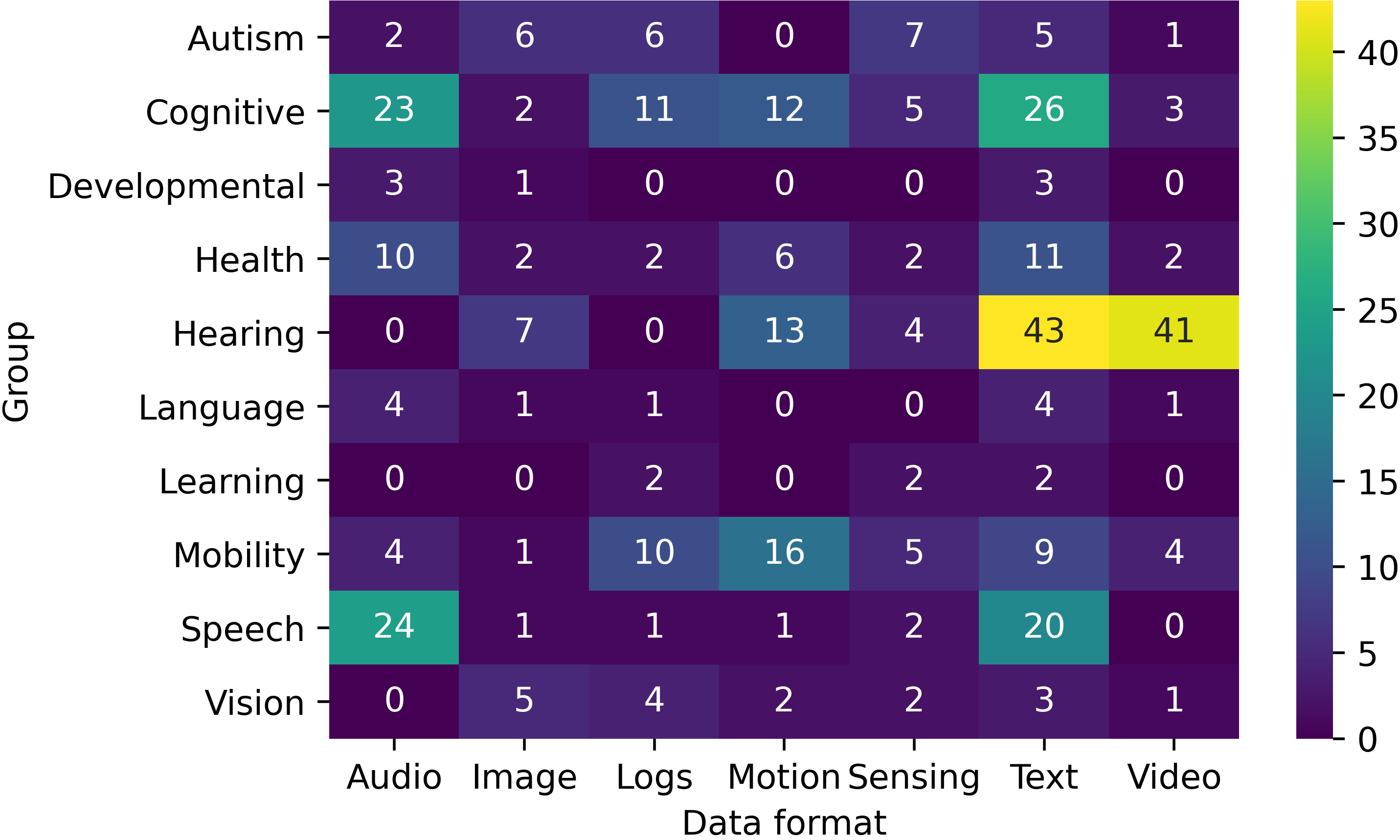}
    \caption{Distribution of data types across communities.}
    \Description[Distribution of groups over the type of data format collected, each row begins with the Group name followed by the number of dataset in the data format: Audio  Image  Logs  Motion  Sensing  Text  Video]{
               Audio  Image  Logs  Motion  Sensing  Text  Video
Group                                                          
Autism         2      6      6     0       7        5     1    
Cognitive      23     2      11    12      5        26    3    
Developmental  3      1      0     0       0        3     0    
Health         10     2      2     6       2        11    2    
Hearing        0      7      0     13      4        43    41   
Language       4      1      1     0       0        4     1    
Learning       0      0      2     0       2        2     0    
Mobility       4      1      10    16      5        9     4    
Speech         24     1      1     1       2        20    0    
Vision         0      5      4     2       2        3     1    }
    ~\label{fig:typeVSdataFormat}
\end{figure}

We describe the different types of data generated across the communities of focus. Figure~\ref{fig:typeVSdataFormat} shows the distribution of data formats across the communities and Figure~\ref{fig:dtypeOverTime} shows how the distribution has evolved over the years. We reviewed all datasets and coded the data formats as Audio, Image, Logs, Motion, Sensing, Text, and Video. A dataset often has multiple formats e.g., the raw data and annotations.

\subsubsection{Audio}
\newcommand\audioTotal{39}
\newcommand\audioSpeech{24}
\newcommand\audioCognitive{23}
\newcommand\audioHealth{10}

This format (in \audioTotal ~datasets) is typically found in the Speech (\audioSpeech) and Cognitive (\audioCognitive) groups, which often overlap as they collect audio samples targeting dysarthric speech~\cite{orozco2016automatic}. We also see audio data in the Health group (\audioHealth) such as speech before and after a neck cancer surgery~\cite{clapham2012nki} or audio data from heart sound signals~\cite{maragoudakis2011mcmc}. Speech tasks range from scripted vowels and sentences~\cite{fougeron2010developing} to story narratives~\cite{wetherell2007narrative}, picture descriptions~\cite{macwhinney2011aphasiabank}, and conversations in pairs~\cite{kempler1987syntactic}, which are produced not only by people with speech impairments but also those with language impairments or intellectual disabilities. The recordings are typically done in an isolated booths, especially in studies that collected vowel or phonation sounds~\cite{fougeron2010developing, aggarwal2018evaluation}. More recently, audio is collected through smartphones and some times in a real-world context, \eg, Parkinson's detection through body sound collected through smartphones~\cite{zhang2019pdvocal}. Audio recordings of speech included English (16), Czech (4), Spanish or Colombian Spanish (4), German (3), French (3), Dutch (2), Korean (1), Italian (1), Taiwanese (1), and Turkish (1). 

\subsubsection{Text}
\newcommand\textTotal{94}
\newcommand\textHearing{43}
\newcommand\textCognitive{26}
\newcommand\textSpeech{20}
\newcommand\textOnly{2}

This is the most common format (in \textTotal ~datasets) as text is often used to supplement and annotate other types of data. For example, gloss labels are used to annotate sign language videos~\cite{efthimiou2007gslc} and phonemic transcriptions are used for spoken utterances of people with cognitive decline~\cite{beltrami2016automatic}. Thus, it is not a surprise to to see this format present in the Hearing (\textHearing), Cognitive (\textCognitive), and Speech (\textSpeech) groups. Only a few (\textOnly) datasets were exclusively text based. They involved Reddit posts related to living with a mental illness (\eg,~\cite{gkotsis2017characterisation}) and corpora related to text readability (\eg~\cite{feng2009automatic}).
Text is also found along side formats such images and logs. For example, VizWiz~\cite{gurari2018vizwiz} includes crowdsourced text responses from sighted people to visual questions asked by blind people.  Another dataset captures cursor tremors along web search queries~\cite{white2019population}.

\subsubsection{Image}
\newcommand\imageTotal{23}
\newcommand\imageAutism{6}
\newcommand\imageHearing{7}
\newcommand\imageVision{5}
This format (in \imageTotal ~datasets) typically includes photos taken by blind individuals, snapshots of signing, and people's faces during interactions. It is most commonly found in the Hearing (\imageHearing), Autism (\imageAutism), and Vision (\imageVision) groups. While sign language data are typically in a video format~\cite{neidle2012new}, images are also used \eg, to capture and analyze individual hand shapes in fingerspelling~\cite{shohieb2015signsworld}; sometimes in combination with depth images~\cite{warchol2019recognition}. Images are also collected in behavioral studies of autistic children often including their face during child-adult~\cite{rehg2014behavioral} and child-robot~\cite{leo2015automatic} interactions.  Photos by people with visual impairments are typically taken from an egocentric viewpoint through smartphone cameras and they tend to include objects to support computer vision technology~\cite{lee2019hands,sosa2017hands}. 
Only one dataset include third-person viewpoint images from a fundus (retinal) camera used for locating blood vessels~\cite{hoover2000locating}.

\begin{figure}
    \includegraphics[width=0.9\linewidth]{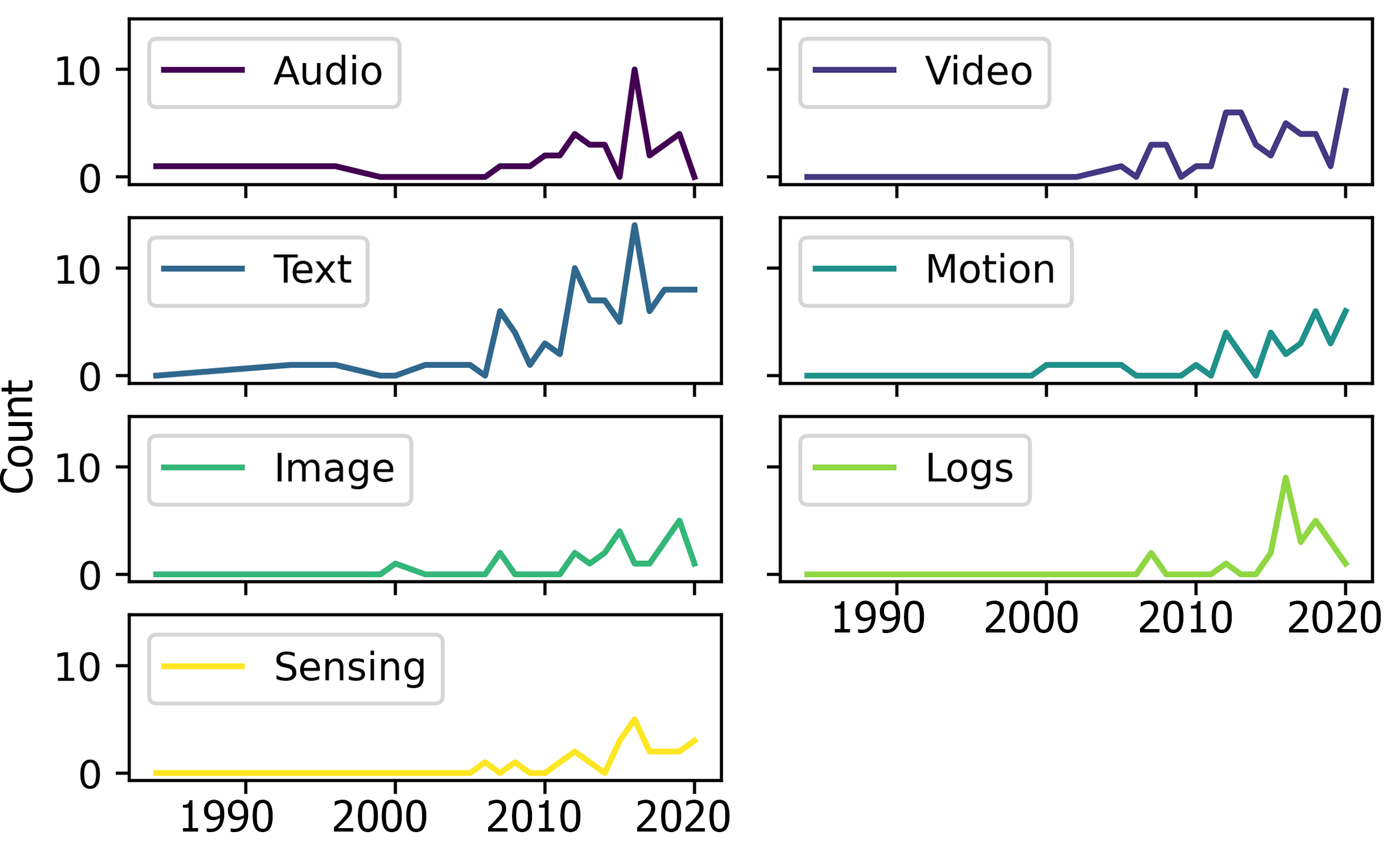}
    \caption{Dataset count over the years across data types.}
    \Description[Number of datasets for each type of data format across the years 1984 to 2020]{
      Audio  Video  Text  Motion  Image  Logs  Sensing
Year                                                  
1984  1      0      0     0       0      0     0      
1993  1      0      1     0       0      0     0      
1996  1      0      1     0       0      0     0      
1999  0      0      0     0       0      0     0      
2000  0      0      0     1       1      0     0      
2002  0      0      1     1       0      0     0      
2005  0      1      1     1       0      0     0      
2006  0      0      0     0       0      0     1      
2007  1      3      6     0       2      2     0      
2008  1      3      4     0       0      0     1      
2009  1      0      1     0       0      0     0      
2010  2      1      3     1       0      0     0      
2011  2      1      2     0       0      0     1      
2012  4      6      10    4       2      1     2      
2013  3      6      7     2       1      0     1      
2014  3      3      7     0       2      0     0      
2015  0      2      5     4       4      2     3      
2016  10     5      14    2       1      9     5      
2017  2      4      6     3       1      3     2      
2018  3      4      8     6       3      5     2      
2019  4      1      8     3       5      3     2      
2020  0      8      8     6       1      1     3      
    }
    ~\label{fig:dtypeOverTime}
\end{figure}

\subsubsection{Sensing}
\newcommand\senseTotal{23}
\newcommand\senseAutism{7}

This format (in \senseTotal ~datasets) involves data collected by sensing instruments that are not easily captured by the motion category such as eye-tracking measurements~\cite{eraslan2019web} or heart sound~\cite{maragoudakis2011mcmc} and EEG~\cite{mamem} signals measured using a medical device. It is typically present in datasets from autism studies (\senseAutism) as they often employ eye tracking with autistic children and adults. Some of these studies try to observe differences in gaze fixations between autistic and non-autistic adults when reading~\cite{yaneva2016corpus}.  Other assess traits by tracking gaze patterns of autistic children when viewing images~\cite{duan2019dataset} or autistic adults when searching information in the web~\cite{eraslan2019web}. Few incorporate multimodal sensing; \eg Rehg~\etal~\cite{rehg2014behavioral}  include wrist-worn sensors recording electrodermal activity and accelerometry. Otherwise, the default is using a stationary eye tracker attached to a display with engineered visual stimuli (\eg, Tobii T120 in~\cite{duan2019dataset}). However, many comment on the limitations of this approach, especially when involving autistic children~\cite{duan2019dataset}. Thus, some attempt to obtain eye contact estimation through egocentric images of the child's face from the experimenter's viewpoint~\cite{rehg2014behavioral}.

\subsubsection{Video}
\newcommand\videoTotal{48}
\newcommand\videoHearing{41}
\newcommand\videoKinect{14}

This format (in \videoTotal ~datasets) captures videos of people signing~\cite{bleicken2016using}, talking~\cite{corrales2016use}, walking~\cite{ahmetovic2018turn}, describing pictures~\cite{depaul2016corpus}, or interacting with technology (\eg, rehabilitation robots~\cite{dolatabadi2017toronto}). The majority (\videoHearing) of these datasets fall in the Hearing group. Typically, they  capture sign language through multiple cameras placed at a fixed location (\eg~~\cite{neidle2012challenges}). 
3D Kinect videos are also common (\videoKinect). These datasets started appearing after 2010, the year Kinect was launched. Looking at the dataset distributions over time in Figures~\ref{fig:communityOverTime} and \ref{fig:dtypeOverTime}, we see that datasets in the Hearing group tend to reflect the availability of research instruments and computational processes enabling fine-grained  analysis of body movements and facial expressions transitioning from video and motion capture to videos only. 
We see a similar trend in Mobility, where datasets typically involve motion sensors but now look at pose estimation in videos(\eg,~\cite{li2018vision}).

\subsubsection{Motion}
\newcommand\motionTotal{35}
\newcommand\motionMobility{15}
\newcommand\motionHearing{14}
\newcommand\motionHealth{5}

This format (in \motionTotal ~datasets) mostly covers data using motion sensors typically embedded in devices such as smartphones to measure hand and walking movements of individuals with mobility impairments affected by neurological disorders such as Parkinson's disease~\cite{bazgir2015neural} or spinal cord injury~\cite{vatavu2019stroke}. Thus, the majority (\motionMobility) of the datasets fall under the Mobility group.  Different approaches are used to collect such data over the last decade; some opt for smartphone-mounted hand gloves to quantify tremor symptoms~\cite{kostikis2015smartphone} others use touchscreen tablets to collect stroke-gesture input~\cite{vatavu2019stroke}. Initially researchers relied on customized motion sensors such as accelerometers and gyroscopes attached in footwear~\cite{klucken2013unbiased} or shoes with force-sensitive resistors measuring foot contact~\cite{hausdorff1997altered}. This format is also present in some (\motionHearing) datasets in the Hearing group, where motion capture gloves and other equipment isused to measure hand and finger positions~\cite{kadous2002temporal, lu2012cuny}. We see some (\motionHealth) datasets also in the Health group capturing pen- and mouse-based interactions of older adults~\cite{moffatt2010addressing} and daily activity movements of people with unipolar or bipolar disorders through wearable sensors~\cite{garcia2018depresjon}.

\subsubsection{Logs}
\newcommand\logsTotal{26}
\newcommand\logsCognitive{11}
\newcommand\logsMobility{9}
\newcommand\logsAutism{6}
\newcommand\logsVision{4}

This format (in \logsTotal ~datasets) typically captures user interactions in mobile applications~\cite{sakar2013collection, kacorri2016supporting}, keyboards~\cite{giancardo2016computer}, and search engines~\cite{white2018detecting, white2019population}. We see this format in the Cognitive (\logsCognitive) and Mobility (\logsMobility) groups with datasets being often cross-listed. The datasets typically include people with Parkinson's. For example, a keystroke logging app, Tappy, was installed on participants' personal computers at home to record key press and release timings for early detection of Parkinson's~\cite{adams2017high}. Logs are also commonly found in the Autism community (\logsAutism) \eg, subjective measures on text complexity by autistic people~\cite{evans2016predicting}, which were often combined with eye tracking measurements for text readability~\cite{yaneva2015accessible, yaneva2016corpus}.

\subsection{Data Sharing Practices}

\newcommand\downloadTotal{52 }
\newcommand\requestTotal{27 }
\newcommand\contactTotal{58 }

\begin{figure}[b]
    \includegraphics[width=1\linewidth]{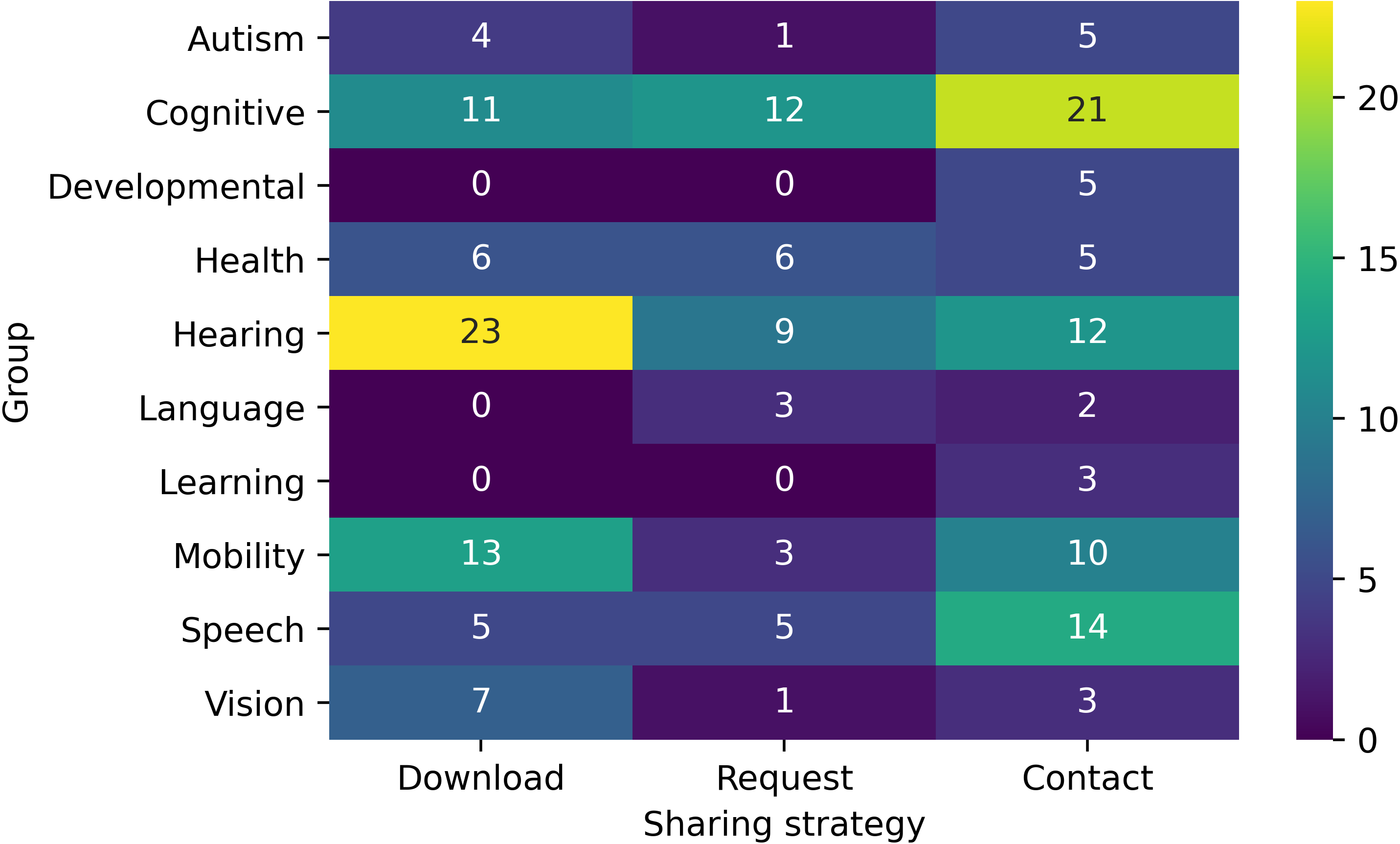}
    \caption{Distribution of sharing strategies across communities.}
    \Description[Distribution of groups over the type of sharing strategy the can be used to access the dataset, each row begins with the Group name followed by the number of dataset that belong to the download strategy: Download  Request  Contact]{
Group                                    
Autism         4         1        5      
Cognitive      11        12       21     
Developmental  0         0        5      
Health         6         6        5      
Hearing        23        9        12     
Language       0         3        2      
Learning       0         0        3      
Mobility       13        3        10     
Speech         5         5        14     
Vision         7         1        3 
    }
    ~\label{fig:typeVSstrategy}
\end{figure}

Figure~\ref{fig:typeVSstrategy} shows how datasets from each community are distributed across strategies that dataset creators employed for sharing or not sharing data. Out of the \datasetsTotal datasets that we analyzed, only \downloadTotal can be downloaded directly (\eg, through a webpage from the dataset creators~\cite{matthes2012dicta}) and \requestTotal are available upon request (\eg, through registration of name, institution, and purpose of the use of data~\cite{leightley2015benchmarking}). The remaining \contactTotal do not include any sharing intent or information; however, we still link to them as we have observed a lack of standardized process for documenting datasets that require guessing or further inquires to understand their intent of sharing. More often, authors choose to provide links or request information for the datasets in some footnote on manuscripts or supplementary materials (\eg, ~\cite{gkotsis2017characterisation}) without specifying data access policy, making it challenging to discover their motivation for sharing data resources. 

\subsubsection{Publicly Available Accessibility Datasets}
\newcommand\projectWeb{26}
\newcommand\Kaggle{4}
\newcommand\UCI{3}
\newcommand\Zenodo{3}
\newcommand\Physio{5}
\newcommand\OrtoLang{3}
\newcommand\Synapse{1}
\newcommand\OSF{1}
\newcommand\downloadHearing{23}
\newcommand\downloadMobility{13}
\newcommand\downloadCognitive{11}
\newcommand\downloadMobCog{8 }
\newcommand\downloadMobCogPhysio{5 }
\newcommand\downloadVision{7}

Datasets in this category can be directly downloaded from personal and project-specific websites (\projectWeb); repositories like Kaggle (\Kaggle), UCI Machine Learning Repository (\UCI), and PhysioNet (\Physio); OrtoLang (\OrtoLang); Zenodo (\Zenodo); Synapse.org (\Synapse); and Open Science Foundation (\OSF). 
This strategy was most commonly found across datasets from the Hearing group (\downloadHearing), mostly related to sign language videos and gloss annotations. The majority of them were shared by computational linguists and computer vision researchers, often having their personal or project site for documentation but the information provided varied in its quality and detail. Dataset creators from ~\cite{neidle2012challenges} documented an overview of participants contributing data, data types and size, and annotation of datasets, as well as a detailed explanation on how the data was collected, processed, and analyzed. Providing such documentation did not seem to be a standard practice; often the creators provide a short description of the dataset on their personal/project web page and cite their related publications for more detail (\eg, ~\cite{kapuscinski2015recognition}). The strategy of direct download was also common among datasets sourced from people who are blind or have low vision (\downloadVision). They were similarly hosted on personal/project websites sharing their photos (\eg,~\cite{gurari2019vizwiz}), touchscreen gestures (\eg,~\cite{vatavu2018impact}), and walking patterns (\eg,~\cite{flores2018weallwalk}).   

We also see the Mobility (\downloadMobility) and Cognitive (\downloadCognitive) groups contributing publicly available datasets \eg, providing keystroke logs collected from users with and without Parkinson's disease~\cite{adams2017high}, where \downloadMobCog of these datasets are cross-listed in these two groups. This community seems to leverage public data repositories more often than personal websites to host their datasets, specifically Kaggle (4) and PhysioNet (4), documenting data purpose and description, along with instructions on how to use the data properly. In a few occasions, they link additional data resources (\eg, analysis tool~\cite{dolatabadi2017toronto}). Relevant acknowledgments and citations are often provided on the repository. Unlike Kaggle that does not specifically target disability data, PhysioNet~\cite{goldberger2000physiobank} is specific to the community for donating and searching biomedical research data and software. It provides a standardized process for documenting datasets, including data access policy and license, and discovery of related publications (\eg,~\cite{hausdorff2000dynamic}).

\subsubsection{Datasets Shared Upon Request}

\newcommand\requestCognitive{12}
\newcommand\requestHearing{9}
\newcommand\requestHealth{6}
\newcommand\requestSpeech{5}
\newcommand\requestLanguage{3}

Datasets in this category can be accessed only upon request through specific procedures. The most common practice we observe is to have a dedicated dataset webpage with a note to contact one of the authors (typically the Project Investigator) given an email address without any further details on eligibility or process. Another practice is to describe the license agreement, the requirements to obtain the data, as well as the types of data that would be shared. This information was either included on the project webpage or included on a dedicated section of the publication where the data were introduced, named \textit{Distribution}. For example, in the BosphorusSign dataset~\cite{camgoz2016bosphorussign} this section reads: \textsl{``The collected corpus will be available to download for academic purposes upon filling a license agreement available from the BosphorusSign website. The provided data will include...
''}

Contributing datasets in this category come from the Cognitive (\requestCognitive) and Hearing (\requestHearing) groups, similar to the representation of these communities in publicly-available datasets though less in number. The contrasting pattern found in datasets shared upon request is the appearance of sharing efforts from the Health (\requestHealth), Speech (\requestSpeech), and Language (\requestLanguage) groups. These datasets were often hosted on TalkBank~\cite{macwhinney2004talkbank}, a data sharing platform for research associated with human communication. It provides separate repositories for different research areas including DementiaBank~\cite{becker1994natural} and AphasiaBank~\cite{macwhinney2011aphasiabank}. Datasets studying communication of patients with Primary Progressive Aphasia (\eg,~\cite{depaul2016corpus}) provide audio and video data of the discourse on AphasiaBank. To access data, the TalkBank system requires membership registration, and clinical data such as those on AphasiaBank are restricted to faculty members or require permission from those who are already a member of the system.

\subsubsection{Non-Shared Datasets}
In our collection, we see that this non-sharing strategy is not unique to a specific community; it is prevalent across data from different user groups including Vision, Hearing, Cognitive, Speech, and Mobility, as well as Autism. More so, we observe that all of our datasets from the Developmental and Learning groups follow this strategy, which encompass the community of what is called ``invisible disabilities,'' disabilities that are less apparent to others and perhaps more sensitive for disclosure. We also observe that children are often involved in these unshared datasets (\eg,~\cite{leo2015automatic,aggarwal2018evaluation}), where parents have to agree to share the videos, images, or audio files that capture their behaviors. Given that the majority of human-computer interaction researchers that work with these populations do not share data~\cite{abbott2019local}, the lack of sharing strategies is not a surprise. Potential factors for not sharing include data sensitivity, participant consent, and re-identification risks~\cite{wacharamanotham2020transparency} as there are increased privacy concerns for accessibility data with the risk of disability disclosure. 

\subsubsection{Datasets Cleared from Ethical Boards}

\newcommand\IRBTotal{49 }
\newcommand\downloadIRB{22}
\newcommand\downloadNonIRB{30}
\newcommand\requestIRB{7}
\newcommand\requestNonIRB{20}

\newcommand\contactIRB{20}
\newcommand\mobilityIRB{19}
\newcommand\cognitiveIRB{25}
\newcommand\hearingIRB{3}

\newcommand\mobilitydownloadIRB{10}
\newcommand\cognitivedownloadIRB{10}
\newcommand\mobcogdownloadIRB{13}
\newcommand\mobcogcontactIRB{13}

Compliance with human subjects research standards and requirements can play an important role in researchers' data sharing practices~\cite{meyer2018practical,barnes2020ethical}. For instance, one cannot share data when the consent form was silent about data sharing or promised that the data would not be shared outside of the research team. Communicating the term ``data'' to participants can also be tricky within the space of the consent form.  We review how many of the papers and resources associated with the datasets in our collection claim that the research has been cleared from institutional review boards (IRBs) or institutions imposing policies and restrictions on the collection and use of research data. 

Out of \datasetsTotal datasets, \IRBTotal reported to be approved by ethical boards, with the following breakdown: Download (\downloadIRB), Request (\requestIRB), and Contact (\contactIRB). Datasets in the Mobility (\mobilityIRB) and Cognitive (\cognitiveIRB) groups, which contribute relatively more publicly available datasets than other, have also higher reporting rate of being cleared from ethical boards. Within these two groups, datasets with mentions of ethical clearance that are downloadable (\mobcogdownloadIRB) constitute of more logs (7) and motion (7) data than respective non-shared datasets (\mobcogcontactIRB) -- here, text (6) and audio (5) are found, often as a combination of speech and transcript data (\eg, 
\cite{rusz2013imprecise}). This reflects the recurring concerns of contributing audio data associated with the risk of speaker identification as mentioned in one of the TalkBank user guidelines~\cite{talkbankirb}.

Interestingly, only a few (\hearingIRB) datasets in the Hearing group, which contributed most of the publicly available datasets in our collection, mentioned ethical clearance in their research work. This could be due to those contributing data not being participants but members of the research team or consultants. We yet note that non-reporting of ethical clearance in publications does not necessarily mean the work was not cleared from ethical boards. NSF-funded projects are required to obtain IRB approval before issuance of an award. The process of projects requesting to be hosted on TalkBank or PhysioNet also involves screening of IRB permission for data sharing. In support, these repositories provide informed consent and data confidentiality guidelines that match the ways they operate (\ie, data restricted to authorized members or open to anyone).

\section{Discussion}

We were able to collect \datasetsTotal datasets that represent populations from different communities in accessibility and aging. Across and within these communities of focus, we observed varying data collection, reporting, and sharing practices that changed over the last three decades with the adoption of technologies and growth of interest, participation, and openness in certain research domains. For example, active contributions from the Hearing group resulted from having a community of researchers from different disciplines often working towards sign language datasets for recognition and translation technologies. Such collaboration was also seen in communities performing clinical studies to collect data for early detection of cognitive or speech impairments. The advances in research instruments, such as Kinect camera for in-depth analysis of body movements or smart devices to collect longitudinal data, have also widened the possibilities for researchers to pursue specific problems in that research space. Along with these research trends, certain communities have already established open data platforms (\eg, TalkBank, PhysioNet) to facilitate data sharing and reuse. However, the majority of datasets are still hosted or announced on personal or project-specific websites. On the other hand, communities that are less represented in data collection and sharing (\eg, Autism, Developmental, Learning) highlight the challenges associated with sharing based on parental consent and child assent as well as involuntary and inaccurate disclosure of visible and invisible disability. 

\textbf{Locating Accessibility Datasets.} Our ``detective'' work highlights that accessibility datasets are difficult to locate and require domain and community knowledge, which can hinder accessible and inclusive AI innovations. This is partially due to inconsistent terminology. The lack of a common language between repositories and publications makes it difficult to find connections and to identify secondary data-reuse cases~\cite{khan2020identifying}. Such issues are more prominent for accessibility datasets, since rigid categorization of disability is difficult~\cite{blaser2020why, berg1992second} and perhaps a questionable task as conditions are vast and fluid~\cite{whittaker2019disability}. There is additional ambiguity in how people describe those contributing data. For example, it is difficult to tell whether ``subjects with cognitive decline'' relate to those ``with early Dementia'' or whether ``signers'' actually indicate Deaf/deaf or hard-of-hearing data contributors. Datasets lack consistent descriptions and require manual screening and, on few occasions guessing or further inquiries, highlighting the importance of a  standardized process for documenting datasets (\eg~\cite{gebru2018datasheets}). More often, links or request information for the datasets are buried in some footnote or a specific section on manuscripts, making it challenging to discover.
Even the Google Dataset search~\cite{google2018datasets} released on September 2018, would leave us wanting for more. To enable broader dataset discovery and transparency, we launched a parallel thread to this work; one involving a data surfacing repository called IncluSet~\cite{kacorri2020incluset}. Of course, assuming that everyone will know about one more repository would defeat the purpose. Thus, we implemented the Google Schema (react-schemaorg), allowing IncluSet to surface accessibility datasets to broader search engines. In IncluSet, researchers don't have to share their data just point to them; anyone coming across an accessibility dataset can point to it and have our team review it.

\textbf{Balancing Risks and Benefits.} In this work we highlight the benefits of creating and sharing accessibility datasets. However, there are many privacy and ethical concerns associated with such practices as people who have distinct data patterns may be more susceptible to data abuse and misuse~\cite{hamidi2018should,treviranus2019value,guo2019toward, abbott2019local}.  Naturally, the risk of deduction~\cite{abbott2019local} can increase when reporting data on smaller populations. For instance, blind participants' age, gender, visual acuity, onset, and mobility aids, typically reported in navigation studies~\cite{kacorri2018environmental}, combined with researchers' location, may reveal their identities to those living in the area. Given that some disability communities can be really small, the effectiveness of privacy-preserving techniques can also be affected~\cite{morris2020ai}, calling for novel approaches (\eg privacy-enhancing distortions on sign language datasets~\cite{bragg2020exploring}). Even when re-identification is not a risk, consent and disclosure can be, as disability status is sensitive. 
We believe that by sharing accessibility datasets we can attract, nurture, and challenge data
scientists and technologists to include people with disabilities and older adults at the forefront of AI innovations. However, the same datasets that are collected to mitigate bias against people with disabilities or to support them through novel AI-infused assistive tech, can be used against them by ``detecting'' their disabilities. This can happen even when disclosure is not voluntary, posing further discrimination risks \eg, for one's healthcare and employment~\cite{whittaker2019disability}.  Thus, we call for better sharing practices as well as technical, legal, and institutional privacy frameworks that are more attuned to concerns from these communities \eg, risks of inaccurate or non-consenting disclosure of a disability.  We hope that researchers utilize our data and insights to reflect and discuss with others the future of data sharing and ownership in accessibility research and other field that are often connected through interdisciplinary efforts. We note that our review is not a call to include underrepresented communities, that we aim to benefit, in models that follow rigid categorization that can pose risks for non voluntary disability disclosure. On the contrary, we are hoping it will help us better understand sharing practices and potential concerns that can feed into the conversations to follow.

\section{Conclusion}
Datasets directly sourced from underrepresented communities such as people with disabilities and older adults can contribute to more inclusive AI applications as well as innovative assistive technologies. However, they are scarce. In this paper, we reflect on the data collection and sharing practices for accessibility datasets across a vast number of disciplines for the past 35 years.  While not an exhaustive search, as dataset search is inherently a challenging task,  our analysis has implications for our fields' position in current practices, and where we should go from here. Specifically, we contribute a deep understanding of the current status of accessibility datasets (1984-2020) in terms of their distribution across communities represented, data collection purpose, and language used to describe the people who contributed with data.  We report trends across data size in terms of the number of people involved referring to both those representing communities of focus and those serving as proxies or control. We explore how data types relate to communities and the purpose of the data collection. More importantly, we identify common data sharing practices and report on clearance from ethical boards.

\section{Acknowledgments}
We thank Sravya Amancherla, Mayanka Jha, Riya Chanduka, and Amnah Mahmood for their contributions in collecting and coding the datasets reported in this paper. We also thank our anonymous reviewers for further strengthening this paper. This work is supported by National Institute on Disability, Independent Living, and Rehabilitation Research (NIDILRR), ACL, HHS (\#90REGE0008).

\bibliographystyle{ACM-Reference-Format}
\bibliography{main}
\end{document}